\documentclass[preprint,floats,aps,12pt,superscriptaddress,nofootinbib,floatfix]{revtex4}
\usepackage{natbib} 
\usepackage{times} 
\usepackage{amssymb,amsmath}
\usepackage{csquotes}
\usepackage{prelim2e}\usepackage[none,bottom]{draftcopy}
\draftcopyName{preprint / }{1.2} %ADAPT TEXT
\ifx\pdfoutput\undefined
\usepackage[usenames,dvips]{color}
\usepackage{graphicx}     % Include figure files
\else
\usepackage[usenames,dvipsnames]{color}
\usepackage{epstopdf}
\usepackage[pdftex]{graphicx}
\usepackage[pdftex]{hyperref}
\hypersetup{%
    pdftitle={Manuscript on long-wave model for drop dynamics on elastic layer} %ADAPT TEXT 
    pdfauthor={et al. and Uwe Thiele},  %ADAPT TEXT 
pdfproducer={lateX},
pdfview=FitV,       % FitH
pdfstartview=FitB,
linkcolor=blue,     % links to same page
citecolor=blue,     % citations
urlcolor=red,      % links to URLs
breaklinks=true,    % links may be split onto 2 lines
colorlinks=true,
citebordercolor=0 0 0,  % color for \cite
filebordercolor=0 0 0,
linkbordercolor=0 0 0,
menubordercolor=0 0 0,
urlbordercolor=0 0 0,
pdfhighlight=/I,
pdfborder=0 0 0,   % no box around links
bookmarksopen=true,
bookmarksnumbered=true
}
\fi
\ifx\pdfoutput\undefined
\DeclareGraphicsExtensions{.eps}
\else
\DeclareGraphicsExtensions{.jpg, .pdf, .tif, .png}
\fi
\graphicspath{{.},{./Figures/}} %$ %ADAPT DIRECTORY
\usepackage[usenames,dvipsnames]{color}
\usepackage[normalem]{ulem}
\definecolor{chcolor}{rgb}{0,0.4,0.6}

\renewcommand{\vec}[1]{\mathbf{#1}}

\begin{document}
%
%----------------------------------------------------------------%
\title{Gradient-dynamics model for liquid drops on elastic substrates}

\author{Christopher Henkel}
%\email{}
%\homepage{}
%\thanks{ORCID ID:}
\affiliation{Institut f\"ur Theoretische Physik, Westf\"alische Wilhelms-Universit\"at M\"unster, Wilhelm-Klemm-Str.\ 9, 48149 M\"unster, Germany}
\author{Jacco H. Snoeijer}
\affiliation{Physics of Fluids Group and J. M. Burgers Centre for Fluid Dynamics, University of Twente, P.O. Box 217, 7500 AE Enschede, The Netherlands}
\author{Uwe Thiele}
\email{u.thiele@uni-muenster.de}
\homepage{http://www.uwethiele.de}
\thanks{ORCID ID: 0000-0001-7989-9271}
\affiliation{Institut f\"ur Theoretische Physik, Westf\"alische Wilhelms-Universit\"at M\"unster, Wilhelm-Klemm-Str.\ 9, 48149 M\"unster, Germany}
\affiliation{Center for Nonlinear Science (CeNoS), Westf{\"a}lische Wilhelms-Universit\"at M\"unster, Corrensstr.\ 2, 48149 M\"unster, Germany}
\affiliation{Center for Multiscale Theory and Computation (CMTC), Westf{\"a}lische Wilhelms-Universit\"at, Corrensstr.\ 40, 48149 M\"unster, Germany}
\begin{abstract}
The wetting of soft elastic substrates exhibits many features that have no counterpart on rigid surfaces. Modelling the detailed elastocapillary interactions is challenging, and has so far been limited to single contact lines or single drops. Here we propose a reduced long-wave model that captures the main qualitative features of statics and dynamics of soft wetting, but which can be applied to ensembles of droplets. The model has the form of a gradient dynamics on an underlying free energy that reflects capillarity, wettability and compressional elasticity. With the model we first recover the double transition in the equilibrium contact angles that occurs when increasing substrate softness from ideally rigid towards very soft (i.e., liquid). Second, the spreading of single drops of partially and completely wetting liquids is considered showing that known dependencies of the dynamic contact angle on contact line velocity are well reproduced. Finally, we go beyond the single droplet picture and consider the coarsening for a two-drop system as well as for a large ensemble of drops. It is shown that the dominant coarsening mode changes with substrate softness in a nontrivial way.
\end{abstract}
%
%\begin{keyword} 
%Sliding drops \sep Heterogeneous substrates \sep Pinning and depinning
%\pacs{
%68.15.+e, % Thin films: Liquid thin films
%47.20.Ky  % Fluid dynamics: Nonlinearity (including bifurcation theory)
%47.55.Dz  % Drops and bubbles 
%68.08.-p  % Liquid-solid interfaces
%}
%\end{keyword} 
%
\maketitle
%
%\received{6.5.2002}
%
%----------------------------------------------------------------%
%
%%%%%%%%%%%%%%%%%%%%%%%%%%%%%%%%%%%%%%%%%%%%%%%%%%%%%%%%%%
%  INTRO
%%%%%%%%%%%%%%%%%%%%%%%%%%%%%%%%%%%%%%%%%%%%%%%%%%%%%%%%%%
%%%%%%%%%%%%%%%%%%%%%%%%%%%%%%%%%%%%%%%%%%%%%%%%%%%%%%%%%%%%%%%%%%%%%%%%%%%%%%%
\section{Introduction} \label{sec:intro}
%%%%%%%%%%%%%%%%%%%%%%%%%%%%%%%%%%%%%%%%%%%%%%%%%%%%%%%%%%%%%%%%%%%%%%%%%%%%%%%
%
The static and dynamic wetting behaviour of simple and complex liquids on various surface types is highly relevant for many aspects of daily life, including cosmetics, cleaning and painting. It is also of great interest for many technical applications like, for example, printing, lubrication, and coating. Related phenomena are studied for more than two hundred years \cite{GennesBrochard-WyartQuere2004,StarovVelardeRadke2007,Bormashenko2017}. Many works published in the past decades consider (de)wetting on smooth homogeneous rigid solid surfaces \cite{Genn1985rmp,BEIM2009rmp}, although more recently other substrate types have also attracted much attention. For instance, in many applications the substrate actually consists of a soft elastic solids \cite{AnSn2020arfm}, viscose liquids \cite{PBMT2005jcp,BCJP2013epje} or viscoelastic materials \cite{KDNR2013sm}. Lubricant-infused surfaces \cite{SXEH2015sm,KKCQ2017sm} are another relevant example. In all these cases, the coupling of the dynamics of the liquid and of the underlying substrate is important. Such phenomena are of particular importance for micro- and mesoscale systems that are dominated by interface phenomena. The latter include adhesion and cohesion forces that, in case of partially wetting liquids, cause droplet formation and also deform the substrate.

A particularly interesting situation arises for the wetting of soft elastic solid substrates \cite{CaGS1996n,CaSh2001l,SJHD2017arcmp,AnSn2020arfm}. This situation is often referred to as ``soft wetting", and gives rise to many wetting phenomena that have no counterpart on a rigid solid. The most prominent feature of soft wetting is the appearance of a wetting ridge: an elastic deformation that is drawn out of the substrate by the vertical pulling force the liquid-gas interface exercises at the three-phase contact line \cite{ExKu1996jcis,JXWD2011prl,BoSD2014sm,PWLL2014nc}. The ridge is responsible for intricate static and dynamical phenomena observed in soft wetting \cite{AnSn2020arfm}.

The fine-structure of the wetting ridge arises from an intricate balance of elastic and capillary forces, with recent literature identifying a central role for the surface tension of the solid~\cite{JXWD2011prl,StDu2012sm,MDSA2012prl,Lima2012epje}. A central question is how the liquid angle, at equilibrium, is different from Young's law for rigid substrates. It turns out that upon increasing the substrate softness, a double-transition is observed in the various angles between interfaces in the contact line region \cite{LWBD2014jfm}. Namely, when increasing softness from low value (i.e., starting with a rigid substrate), first a strong increase in the size of the wetting ridge is observed while the droplet shape remains approximately constant and the substrate surface in the inner region remains smooth. Then, when the softness decreases several orders of magnitude, a strong increase of the substrate depression under the droplet occurs a constant opening angle of the wetting ridge. The solid angle is governed by a balance of surface tensions \cite{StDu2012sm,MDSA2012prl,Lima2012epje}, analogously to the Neumann's law known for floating lenses. In comparison to liquids, however, the surface tension of solids brings in an additional complexity owing to its possible strain-dependence (i.e., the Shuttleworth effect)  \cite{AnSn2016el,XJBS2017nc,XuSD2018sm,STSR2018nc,PAKZ2020prx}.
The wetting ridge further noticeably influences the dynamics of moving contact lines and droplet interactions. Depending on the substrate layer thickness and softness the wetting ridge may develop local depressions on both sides which counteract coarsening by droplet translation and can even result in an effective repelling force between them -- also referred to as the ``inverted Cheerios effect'' \cite{KPLW2016pnasusa}.
Another consequence is the so called ``viscoelastic braking", describing the decreasing speed of a moving contact line with increasing softness of the substrate, due to increased dissipation within the latter. An effect that was already investigated by Carr{\'e} and Shanahan \cite{CaGS1996n,CaSh2001l} and Long, Ajdari \& Leibler \cite{LoAL1996lb}, followed up in recent years in Refs.~\cite{ChKu2020sm,KDGP2015nc,GKAS2020sm}. New phenomena were discovered, such as intricate dependencies of sliding velocity on substrate thickness \cite{ZDNL2018pnasusa} and stretching \cite{SXHB2021prl}, as well as effects of free polymer chains that can be present in the substrate \cite{HACN2017sm}. Besides slowing down the contact line, the wetting ridge can also be the source of complex periodic \textit{stick-slip} motion \cite{KDNR2013sm,KDGP2015nc,GASK2018prl}.

From the modelling perspective, various approaches to soft wetting have been proposed. Most analytical results have been obtained using linear elasticity theory \cite{LoAL1996lb,JXWD2011prl,StDu2012sm,Lima2012epje,LWBD2014jfm,BoSD2014sm,KDGP2015nc}. In these approaches the surface tension forces are treated as tractions that are imposed as boundary conditions at the free surface. This approach has now been extended to large deformations, including the possibility of the Shuttleworth effect  \cite{WLJH2018sm,PAKZ2020prx}. Other studies employed an approach where the tractions are not imposed by hand, but computed via a Derjaguin (or disjoining) pressure \cite{StVe2009jpm}) -- this is a hybrid approach coupling elasticity to a thin-film static equation  \cite{Whit2003jcis} or dynamic evolution equation for the liquid layer~\cite{MaGK2005jcis,GLBG2017csaea,ChKu2020sm}. Only recently, 
dynamic calculations coupling full hydrodynamic and elasticity models have become available \cite{AlMo2021ijnme, AlAu2021ijnme, RoBr2019mmmas, BRSZ2017}. 
On the other end of the spectrum, some studies employ molecular dynamics simulations \cite{WeAS2013sm,LCWD2018l}, exploring the elastocapillary interactions at the nanoscale.

None of the reviewed modelling approaches lends itself easily to large-scale simulations considering, e.g., the coarsening behaviour of large droplet ensembles on soft substrates. Also the incorporation of additional effects, e.g., the Shuttleworth effect or forces due to thermal or solutal Marangoni effects or condensation/evaporation, is a major effort. In the present work we present a simple qualitative model that captures main features of the coupled dynamics of elastic substrate and liquid drop or film. It is developed as a gradient dynamics model \cite{ThAP2016prf,Thie2018csa} on a simplified free energy that in its simplest form captures compressional elasticity but can be extended to include effects such as those mentioned above. Such a model may then be employed in large-scale simulations similar to Refs.~\cite{WTEG2017prl,EnTh2019el} in the case of droplet ensembles on smooth rigid substrates.

The collective behaviour of such ensembles, e.g., their coarsening behaviour, is of large interest in practical contexts like, for instance, condensation or inkjet printing. Similar coarsening processes (Ostwald ripening) occur for emulsion droplets \cite{Tayl1998acis}, quantum dots \cite{VeGY2001s}, or crystalline nanoparticles \cite{RNRH2014n}. In all cases, the mean cluster or drop size and their mean distance continuously grow following power laws. For simple nonvolatile liquids on horizontal homogeneous substrates is experimentally well studied \cite{ABNP2002jfm,LiGr2003l,BLHV2012prl,BDHE2018prl} and theoretically well understood through simulations and asymptotic considerations \cite{GrWi2009pd,GORS2009ejam,Kita2014ejam} based on thin-film models \cite{Wite2020am}. As explained more in detail below in section~\ref{sec:coarse} one distinguishes two coarsening modes - Ostwald ripening and migration - with the former often seen as dominant. However, for the dewetting process of a liquid film on a rigid substrate, Refs.~\cite{GlWi2005pd,GORS2009ejam} find that, in contrast, there exist extended parameter regions where coarsening by migration dominates. We employ the developed simple model to investigate how substrate softness influences the dominant coarsening mode.

Here, the mesoscopic thin-film model is presented in the subsequent section~\ref{sec:model} where we also discuss how the quantities of the model are related to macroscopic interface tensions and, in consequence, how the classical Young and Neumann laws result as limiting cases. In section~\ref{sec:steady} the model is employed to study the double transition of steady drops in its dependence on drop size, section~\ref{sec:spreading} investigates the spreading dynamics of a single drop in dependence of substrate softness for partially and completely wetting liquids, and section~\ref{sec:coarse} considers the coarsening behaviour of two drops and large drop ensembles, again in dependence of substrate softness. The various results are discussed in the context of literature results. 
Section~\ref{sec:conc} concludes and gives an outlook.

%%%%%%%%%%%%%%%%%%%%%%%%%%%%%%%%%%%%%%%%%%%%%%%%%%%%%%%%%%%%%%%%%%%%%%%%%%%%%%%
\section{Modelling approach}
\label{sec:model}
\subsection{Governing equations}
\label{sec:goveq}
%%%%%%%%%%%%%%%%%%%%%%%%%%%%%%%%%%%%%%%%%%%%%%%%%%%%%%%%%%%%%%%%%%%%%%%%%%%%%%%
\begin{figure}[tbh]
		\includegraphics[scale=1.]{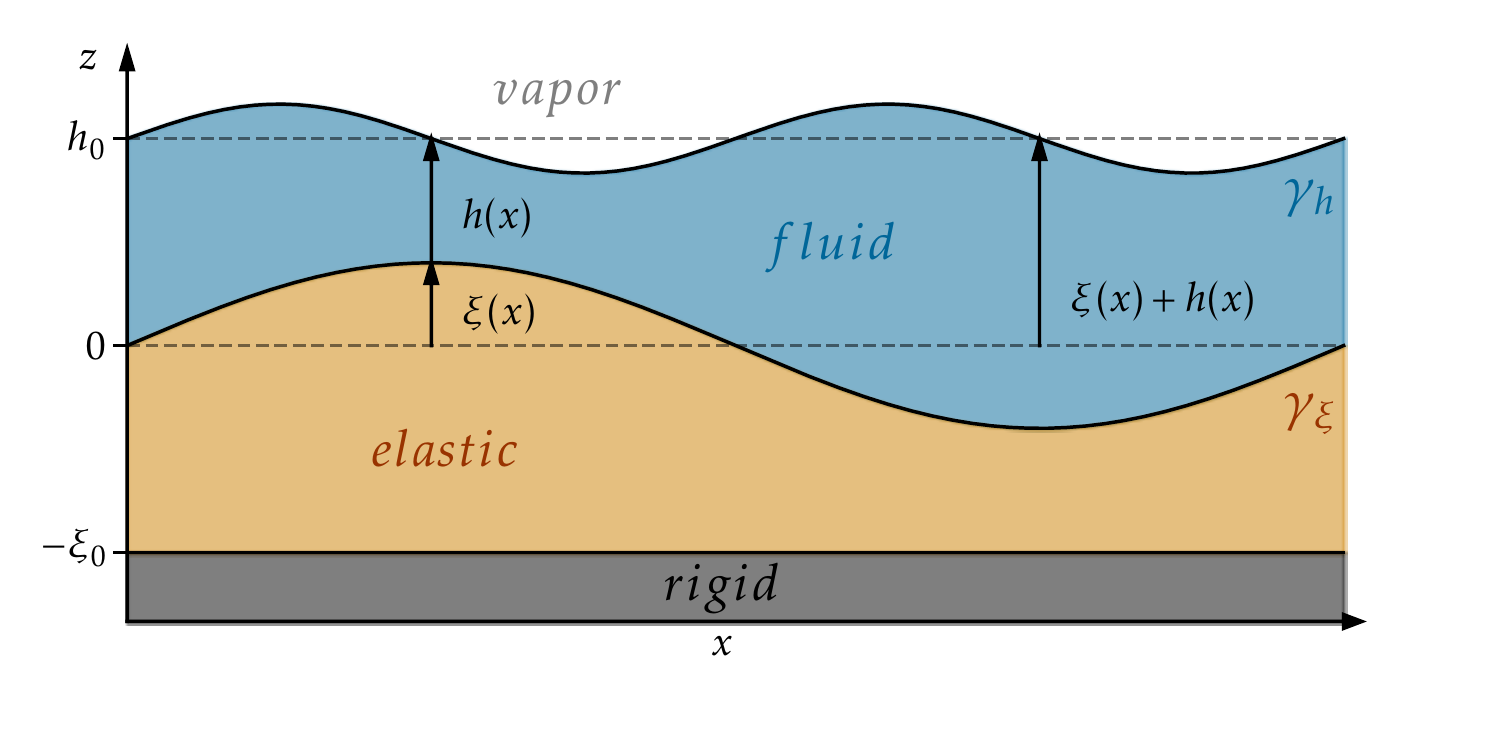}
	\caption{The sketch shows the geometry of the considered two-layer system consisting of a liquid layer on an elastic layer. The description is based on the liquid layer thickness $h(\vec{x},t)$ and the vertical deviation $\xi(\vec{x},t)$ of the elastic-liquid interface from a uniform reference height. The mean thicknesses of the layers are given by $h_0$ and $\xi_0$, respectively. The liquid-gas and the elastic-liquid interface have constant interface energies $\gamma_h$ and $\gamma_\xi$, respectively.}
	\label{fig:sketch}
      \end{figure}
     The geometry of the problem is sketched in Fig.~\ref{fig:sketch}, where we define the liquid layer thickness $h(\vec{x},t)$, and the deviation of the elastic layer thickness from a uniform reference height $\xi(\vec{x},t)$. 
     We employ a long-wave model \cite{OrDB1997rmp,CrMa2009rmp} consisting of a standard thin-film equation for $h(\vec{x},t)$ \cite{Thie2010jpcm} and a  nonconserved Allen-Cahn-type relaxational dynamics for $\xi(\vec{x},t)$.    
     In dimensional form we have
\begin{eqnarray}
\partial_t h &=& \nabla\cdot\left[\frac{h^3}{3\eta}\nabla p_h\right],\nonumber\\
\partial_t\xi &=& -\frac{1}{\zeta} \, p_\xi. \label{eq:dynamics}
\end{eqnarray}
Here, $p_h$ and $p_\xi$ are the pressures at the two interfaces that we will express in terms of functional derivatives. We further introduced the viscosity of the liquid $\eta$, and an ``elastic friction constant'' $\zeta$ that governs the relaxation of the elastic layer. We remark that the dynamics of the elastic layer is a spatially continuous equivalent of a compressible Kelvin-Voigt-type model, as will be further discussed below.

The system of governing equations is of gradient dynamics form \cite{Thie2018csa}, i.e., the pressures $p_h$ and $p_\xi$ are obtained as variations of an underlying free energy functional
\begin{equation}
\mathcal{F}[h,\xi] = \mathcal{F}_\mathrm{cap}[h,\xi] +  \mathcal{F}_\mathrm{wet}[h] +  \mathcal{F}_\mathrm{el}[\xi]
\label{eq:energy}
\end{equation}
with respect to $h$ and $\xi$, respectively. The energy~Eq.~(\ref{eq:energy}) features contributions due to capillarity of both interfaces, wettability of the liquid on the elastic layer, and elasticity. 
The first contribution consists of the interface energies in long-wave approximation, e.g., the area element $d^2s=\sqrt{1+|\nabla h|^2} \,d^2x$ of the liquid-gas interface is approximated by $d^2s\approx \left(1+\frac{1}{2}|\nabla h|^2\right) d^2x$. Up to a constant, one then has
\begin{equation}
\mathcal{F}_\mathrm{cap}[h,\xi] = \int_\Omega \left[\frac{\gamma_h}{2}|\nabla (\xi+h)|^2+\frac{\gamma_\xi}{2}|\nabla\xi|^2\right] d^2x
\label{eq:energy-cap}
\end{equation}
where $\Omega$ is the considered domain, and $\gamma_h$ and $\gamma_\xi$ are the liquid-gas and the elastic-liquid interface energies per surface area, respectively. Partial wettability of the liquid on the elastic layer is incorporated into our mesoscopic model via
\begin{equation}
\mathcal{F}_\mathrm{wet}[h] = \int_\Omega f(h) d^2x
\label{eq:energy-wet}
\end{equation}
where the local wetting energy $f(h)$ is related to the Derjaguin (or disjoining) pressure $\Pi(h)=-df/dh$ \cite{Genn1985rmp,StVe2009jpm}. Here we use a simple form combining destabilising long-range van der Waals interactions and stabilising short-range adhesion forces, namely,
\begin{equation}
f(h) = \frac{A_H}{2h^2}\left[\frac{2}{5}\left(\frac{h_\mathrm{a}}{h}\right)^3-1\right].
\label{eq:energy-wet-f}
\end{equation}
In this way, even macroscopically 'dry' spots are covered by a thin adsorption (or precursor) layer of thickness $h_\mathrm{a}$ (a molecular scale) \cite{PODC2012jpm}. The Hamaker constant $A_H>0$ is related to the (small) equilibrium contact angle $\theta_\mathrm{eq}$ by $A_H=\frac{5}{3}\gamma_h h_\mathrm{a}^2 \theta_\mathrm{eq}^2$. As a result the effective elastic-gas interface energy is
$\gamma_{sv}=\gamma_h+\gamma_\xi+f(h_\mathrm{a})=\gamma_h+\gamma_\xi- \frac{\gamma_h}{2} \theta_\mathrm{eq}^2$. In the limit of a rigid flat lower layer, this directly corresponds to the Young-Laplace law in long-wave approximation ($\cos \theta_\mathrm{eq}\approx 1 - \frac{1}{2} \theta_\mathrm{eq}^2$). 

Finally, we need to specify the elastic energy, which in linear elasticity theory can be expressed as a quadratic non-local functional of $\xi(\mathbf x,t)$. Here we approximate the energy\footnote{As given e.g., in Ref.~\cite{LoAL1996lb} in Fourier space.}
by a local form, as 
\begin{equation}
\mathcal{F}_\mathrm{el} =\frac{\kappa_v}{2} \int_\Omega \xi(\vec{x},t) \int_\Omega Q(\vec{x}-\vec{x}')\xi(\vec{x}',t) d^2x' d^2x \approx \frac{\kappa_v}{2} \int_\Omega \xi^2 d^2x.
\label{eq:energy-elast}
\end{equation}
This gives a simple continuous spring potential $\frac{\kappa_v}{2}\xi^2$,  where $\kappa_v$ corresponds to a spring constant per unit area. Such a free energy corresponds to the so-called Winkler foundation model \cite{Johnson1987}, used also in problems of soft lubrication \cite{SkMa2004prl,SaMa2015jfm}. In Appendix~\ref{sec:app-Gkappav} we show how the above model can in principle be related to an elastic modulus $G$.

Varying the functional~(\ref{eq:energy}) one obtains the pressures $p_h=\frac{\delta\mathcal{F}}{\delta h}$ and $p_\xi=\frac{\delta\mathcal{F}}{\delta \xi}$. After nondimensionalisation the governing equations \eqref{eq:dynamics} become
\begin{eqnarray}
	\partial_t h &=& -\nabla\cdot\left(h^3\nabla\left[\Delta(h+\xi)+\Pi(h)\right]\right), \label{eq:dynamics-nondim-h}\\
	\partial_t \xi &=& \frac{1}{\tau}\left[\Delta(h+\xi)+\sigma\Delta\xi-\kappa\xi\right]. \label{eq:dynamics-nondim-xi}
\end{eqnarray}
where now all quantities are nondimensional.
Here, the used rescaling for space and time is based on quantities of the liquid layer: vertical length scale $h_\mathrm{a}$, horizontal  length scale $L=\sqrt{3/5} h_\mathrm{a}/\theta_\mathrm{eq}$, time scale $\tau_h=27 h_\mathrm{a}\eta/25 \gamma_h \theta_\mathrm{eq}^4$. Note that the employed long-wave approximation is valid if the ratio of the introduced vertical and horizontal length scales $\epsilon=h_a/L=\sqrt{5/3}\theta_\mathrm{eq}$ is small. The dynamical system is governed by three dimensionless parameters that are all contained in the time-evolution equation for $\xi$:
\begin{equation}
\sigma=\frac{\gamma_\xi}{\gamma_h}, 
\quad \tau= \frac{5}{9} \frac{\zeta h_\mathrm{a} \theta_\mathrm{eq}^2}{\eta}, 
\quad \kappa=s^{-1}= \frac{3}{5} \frac{\kappa_v h_\mathrm{a}^2}{\gamma_h \theta_\mathrm{eq}^2}.
\end{equation}
The parameter $\sigma$ is the ratio of interfacial energies, $\tau$ reflects the ratio of viscous and viscoelastic dissipation, while $\kappa$ is the dimensionless stiffness (or the inverse softness $s^{-1}$). Below we discuss how these dimensionless quantities can be interpreted in a macroscopic framework. Finally, Eq.~(\ref{eq:dynamics-nondim-h}) contains the dimensionless Derjaguin pressure $\Pi(h)=h^{-6}-h^{-3}$. In the scaled long-wave units, the equilibrium angle on a rigid substrate is $\sqrt{3/5}$.

The presented model has a natural thermodynamic structure. The change in the free energy is given by
\begin{align}
\frac{d\mathcal{F}}{dt}=&\int_\Omega \frac{\delta\mathcal{F}}{\delta h}\frac{\partial h}{\partial t} + \frac{\delta\mathcal{F}}{\delta \xi}\frac{\partial \xi}{\partial t}~d^2x \label{eq:diss1}\\
%=&\int_\Omega \nabla\left[h^3\nabla\frac{\delta\mathcal{F}}{\delta h}\right]\cdot\frac{\delta\mathcal{F}}{\delta h}~dx -\int_\Omega \tau \left(\frac{\delta\mathcal{F}}{\delta\xi}\right)^2~dx \label{eq:diss2}\\
=&-\int_\Omega h^3\left|\nabla\frac{\delta\mathcal{F}}{\delta h}\right|^2~d^2x -\int_\Omega \frac{1}{\tau} \left(\frac{\delta\mathcal{F}}{\delta\xi}\right)^2~d^2x~<~0 \label{eq:diss3}
\end{align}
where from (\ref{eq:diss1}) to (\ref{eq:diss3}) we used Eqs.~(\ref{eq:dynamics-nondim-h}) and (\ref{eq:dynamics-nondim-xi}) as well as integration by parts (in the first term). It is always negative and corresponds to the negative of the total dissipation $\mathcal{D}$, i.e., $\mathcal{D}=-d\mathcal{F}/dt$. A useful feature of Eq.~(\ref{eq:diss3}) is that it allows one to individually measure the contributions coming from the liquid and elastic layer, respectively and to resolve them spatially using the dissipation density $\Gamma(\vec{x},t)$, i.e. $\mathcal{D}=\mathcal{D}_h+\mathcal{D}_\xi=\int_\Omega\Gamma_h~d^2x+\int_\Omega \Gamma_\xi~d^2x = \int_\Omega \Gamma~d^2x$.

When determining steady states, $\partial_t h = \partial_t \xi = 0$ and one may integrate Eq.~(\ref{eq:dynamics-nondim-h}) twice to obtain 
\begin{eqnarray}
	0 &=& \Delta(h+\xi)+\Pi(h)+\mu, \label{eq:statics-nondim-h}\\
	0 &=& \frac{1}{\tau}\left[\Delta(h+\xi)+\sigma\Delta\xi-\kappa\xi\right]. \label{eq:statics-nondim-xi}
\end{eqnarray}
where $\mu$ is a Lagrange multiplier ensuring mass conservation
\begin{equation}\label{eq:mass_cons}
\int_\Omega (h-h_0)~d^2x = 0.
\end{equation}
where $h_0$ is the mean thicknesses of the liquid. The latter is related to the drop volume $V_\mathrm{d}$ above the adsorption layer by $V_\mathrm{d} \approx (h_0-1)D$ where $D$ is the area of the domain. In the following we only consider the physically two-dimensional situation, i.e., the substrate is one-dimensional (1d). All given equations remain valid with $\nabla\to\partial_x$, $\Delta\to\partial_{xx}$, $d^2x \to dx$, and $\vec{x}\to x$. Then, the domain size $D$ is a length.

\subsection{Relation to macroscopic theory}

\begin{figure}[tbh]
	\includegraphics[scale=1.]{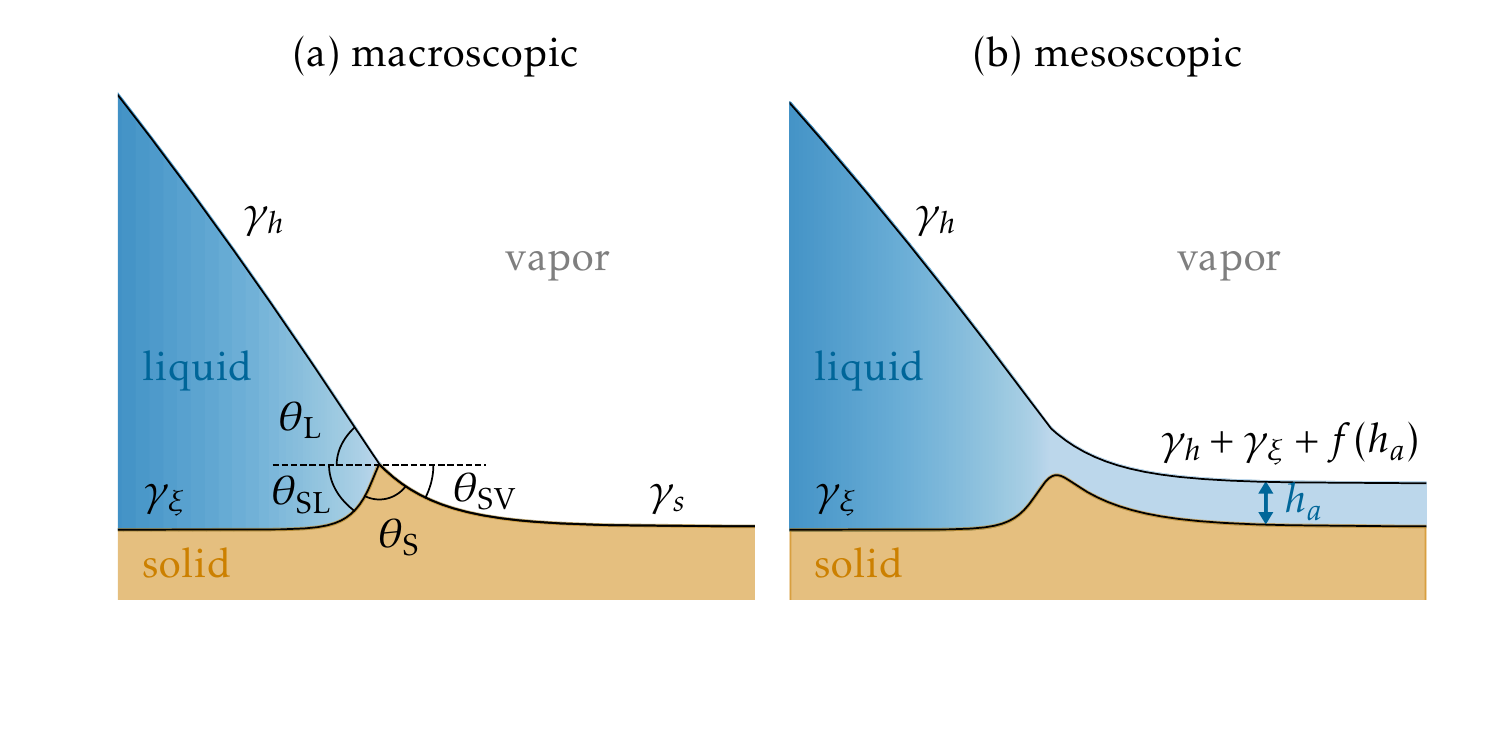}
	\caption{Sketch of the contact line region from two different perspectives: (a) macroscopic, for which there are sharply defined contact angles and contact point, (b) mesoscopic, where we observe an adsorption layer of thickness $h_a$ on the macroscopically dry substrate and its macroscopic interface energy $\gamma_s$ is expressed using the wetting potential $f(h)$. The macroscopic contact point becomes a contact region and the corresponding point force becomes a smoothened-out pressure.}
	\label{fig:ContactRegion}
\end{figure}

It is instructive to interpret the proposed mesoscopic model in the context of the usual macroscopic description. Both perspectives are sketched in Fig.~\ref{fig:ContactRegion}. In the literature of elastic wetting, it is common to express the deformability in terms of the ``elastocapillary length" $\ell_{\rm ec}$, dictating the typical elastic deformation induced by surface tension. In the present model, such a length scale can be defined as 
\begin{equation}\label{eq:lengths}
\ell_\mathrm{ec}=\sqrt{\frac{\gamma_\xi}{\kappa_v}}, \quad  
\Rightarrow \quad \tilde \ell_{\rm ec} = \frac{\ell_{ec}}{L}  = \sqrt{\frac{\sigma}{\kappa}}  = \sqrt{\sigma s}.
\end{equation}
In the second step we defined the dimensionless ratio $\ell_{\rm ec}/L$, based on the horizontal scale. Hence, the softness parameter $s$ can be expressed in terms of the ratio of the typical elastic deformation of the substrate to typical horizontal scale. The latter is proportional to the microscopic thickness of the adsorption layer divided by the small equilibrium contact angle $\theta_\mathrm{eq} $.

A central feature of the macroscopic description is that it exhibits well-defined contact angles, as indicated in Fig.~\ref{fig:ContactRegion}~(a). Given that the drops are typically very large compared to the scale $h_a$, these angles can also be observed in the mesoscopic model, at intermediate distances from the contact line (much larger than $h_a$, much smaller than the drop size). If in addition we consider $\kappa=0$, for which there is no elasticity and the substrate is a liquid layer, one can express the contact angles directly in the interfacial energies -- using Neumann's law. In the non-dimensional long-wave limit the Neumann conditions read
\begin{subequations}\label{eq:neumann}
	\begin{align}
	&\text{horizontal:}&\theta_\text{L}-\sigma\theta_\text{SL} = \left[1+\sigma+\epsilon^2f(h_a)\right]\theta_\text{SV},\\
	&\text{vertical:}&\theta_\text{L}^2+\sigma\theta_\text{SL}^2 +2f(h_a) = \left[1+\sigma+\epsilon^2f(h_a)\right]\theta_\text{SV}^2.
	\end{align}
\end{subequations}
In the absence of any elasticity, the liquid-liquid system approaches the shape of a liquid lens, i.e. as $\kappa$ approaches zero the angle $\theta_\text{SV}$ approaches zero as well. Neglecting this angle from the Neumann conditions \eqref{eq:neumann}, i.e. setting the right hand sides in Eqs.~(\ref{eq:neumann}) to zero, the remaining angles may be expressed like
\begin{subequations}\label{eq:neumann_lens}
	\begin{align}
	&\text{horizontal:}&\theta_\text{L}=\sigma\theta_\text{SL},\\
	&\text{vertical:}&\theta_\text{SL} = \sqrt{\frac{-2f(h_a)}{\sigma(1+\sigma)}}.
	\end{align}
\end{subequations}
These relations give a well-defined macroscopic meaning to the dimensionless parameter $\sigma$ (ratio of surface energies) and the wetting energy of the adsorption layer $f(h_a)$. We remind that in dimensionless units $f(h_a) = - 3/10$, which governs the liquid angle for the rigid substrate ($\kappa=\infty$), i.e., in long-wave scaling $\tilde{\theta}_\mathrm{eq}=\sqrt{-2f(h_a)}$. Indeed, we will verify below that the ``liquid" and ``rigid" limits are correctly recovered by the mesoscopic model.

Finally, let us comment on the final dimensionless number $\tau$, which expresses a ratio of timescales. The substrate equation (\ref{eq:dynamics-nondim-xi}) contains a stiffness $\kappa_v$ and dissipative constant $\zeta$. As is common for the Kelvin-Voigt model, a timescale is obtained by taking the ratio $\zeta/\kappa_v$. Exciting the solid at slower timescales gives a nearly elastic response, while at faster timescales the layer is highly dissipative. We thus introduce the substrate's viscoelastic timescale $\tau_\xi$, defined as
\begin{equation}\label{eq:times}
\tau_\xi= \frac{\zeta}{\kappa_v}, \quad  
\Rightarrow \quad \tilde{\tau} = \frac{\tau_\xi}{\tau_h}  = s \tau = \frac{\tau \ell_{\rm ec}^2}{\sigma}. 
\end{equation}
In the second step we expressed the ratio $\tau_\xi/\tau_h$, comparing the viscoelastic timescale to that of the viscous fluid. If one would chose to scale the model on the \emph{substrate} properties, most notably using $\ell_{\rm ec}$ and $\tau_{\xi}$,  Eq.~(\ref{eq:dynamics-nondim-xi}) took the form 
$$\partial_t \xi = \frac{1}{\tilde\tau}\left[\frac{\tilde\ell_\mathrm{ec}^2}{\sigma}\Delta(h+\xi)+\tilde\ell_\mathrm{ec}^2\Delta\xi-\xi \right].$$
In what follows, we stick to the original formulation, using $\tau,\sigma,\kappa$ as dimensionless numbers, obtained when scaling the variables on the fluid properties. For a physical interpretation of the results, however, we will make use of the connections (\ref{eq:lengths}) and (\ref{eq:times}).

\subsection{Numerical approach}
\label{sec:num}
%%%%%%%%%%%%%%%%%%%%%%%%%%%%%%%%%%%%%%%%%%%%%%%%%%%%%%%%%%%%%%%%%%%%%%%%%%%%%%%
In the following usage of the developed model in a number of static and dynamic situations we use two numerical techniques, namely the continuation of steady states and direct time simulations. For both methods the one-dimensional spatial domain $\Omega$ of size $D$ is discretized into a normally nonuniform adaptive grid with periodic or Neumann boundary conditions.

Steady solutions are obtained by solving the second order Eqs.~(\ref{eq:statics-nondim-h}) and~(\ref{eq:statics-nondim-xi}) by pseudo-arclength continuation \cite{KrauskopfOsingaGalan-Vioque2007,DWCD2014ccp,EGUW2019springer} employing the packages \texttt{auto-07p} \cite{DoKK1991ijbcb} and \texttt{PDE2path} \cite{UeWR2014nmma}. They are frequently used for thin-film descriptions of layers and drops of simple or complex liquids on solid or liquid substrates \cite{PBMT2005jcp,Thie2010jpcm,EWGT2016prf,TSJT2020pre}. Continuation is started at an analytically or numerically known steady state, e.g., a flat film with a small harmonic modulation of a wavelength given by a linear analysis as critical value for a surface instability. Then the steady state is followed in parameter space using, e.g., softness as control parameter, and fixing the drop volume via a Lagrange multiplier (here $\mu$). An individual continuation step combines a tangent predictor and Newton correction steps \cite{EGUW2019springer,Thiele2021lectureCont}. 

\begin{figure}[tbh]
	\includegraphics[width=0.7\hsize]{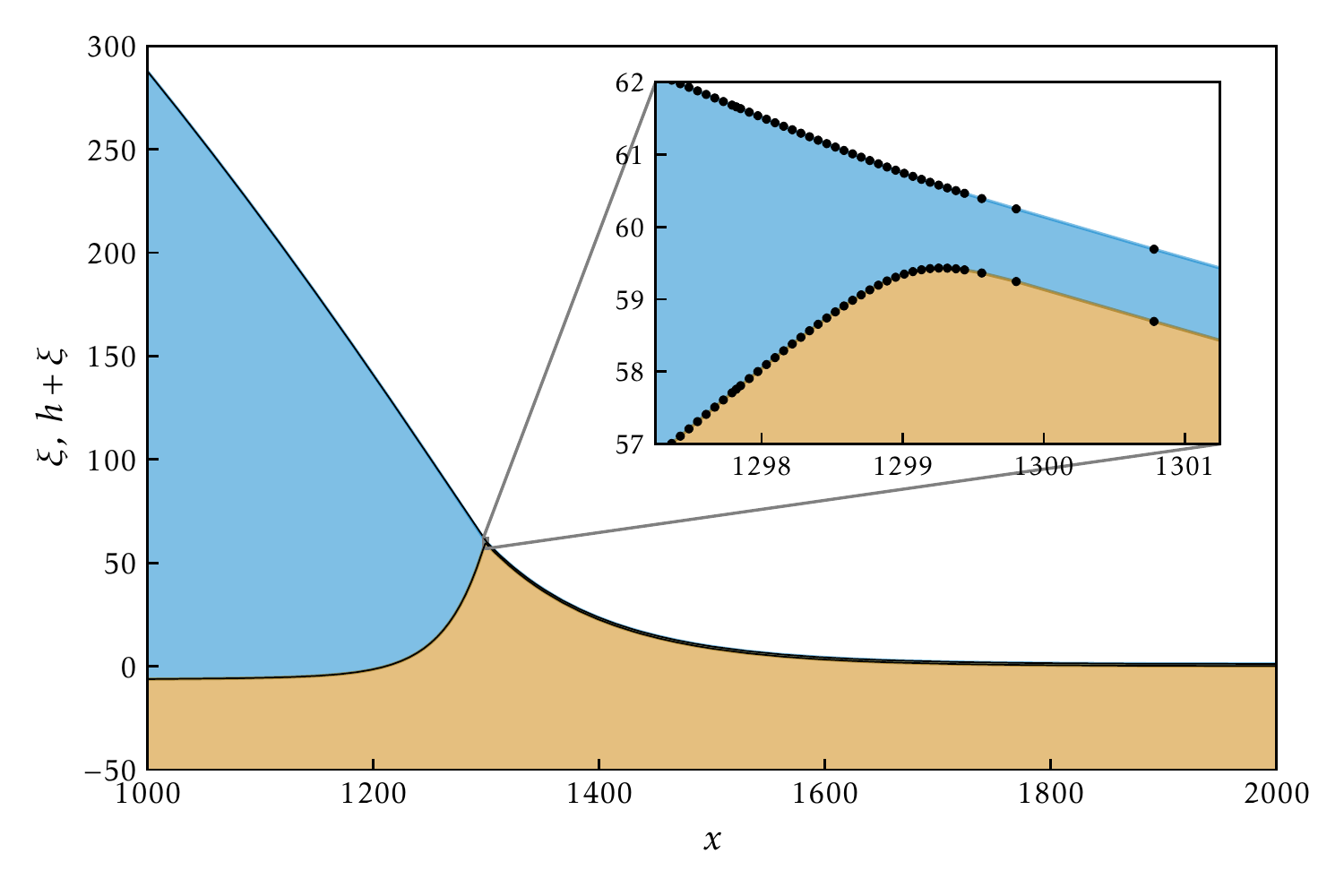}
	\caption{Typical drop and substrate profile in the contact line region. The inset strongly magnifies the tip of the wetting ridge illustrating that it is indeed rounded although in the main panel it appears to be quite pointed. The black dots in the inset show an example of the nonuniform adaptive numerical grid.}
	\label{fig:CRZoom}
\end{figure}

For the direct time simulations of Eqs.~(\ref{eq:dynamics-nondim-h}) and~(\ref{eq:dynamics-nondim-xi}) the FEM-based software package \textsc{oomph-lib} \cite{HeHa2006} is employed. An adaptive time stepping is used based on a backward differentiation method of order 2 (BDF2) from which the next state is obtained via a Newton procedure. The efficient adaptive time stepping and mesh refinement routines allow for a treatment of even very large systems. Fig.~\ref{fig:CRZoom} illustrates that even quite sharp corners like the tip of a wetting ridge are properly resolved.

In the following sections we employ the model to study the double transition of steady drops (section~\ref{sec:steady}), the spreading dynamics of a single drop (section~\ref{sec:spreading}) and the drop coarsening behaviour (section~\ref{sec:coarse}). The results are compared to findings in the literature to evaluate the validity of the developed model.

%%%%%%%%%%%%%%%%%%%%%%%%%%%%%%%%%%%%%%%%%%%%%%%%%%%%%%%%%%%%%%%%%%%%%%%%%%%%%%%
\section{The double transition for static drops} \label{sec:steady}
%%%%%%%%%%%%%%%%%%%%%%%%%%%%%%%%%%%%%%%%%%%%%%%%%%%%%%%%%%%%%%%%%%%%%%%%%%%%%%%

\begin{figure}[tbh]
		\includegraphics[width=0.8\hsize]{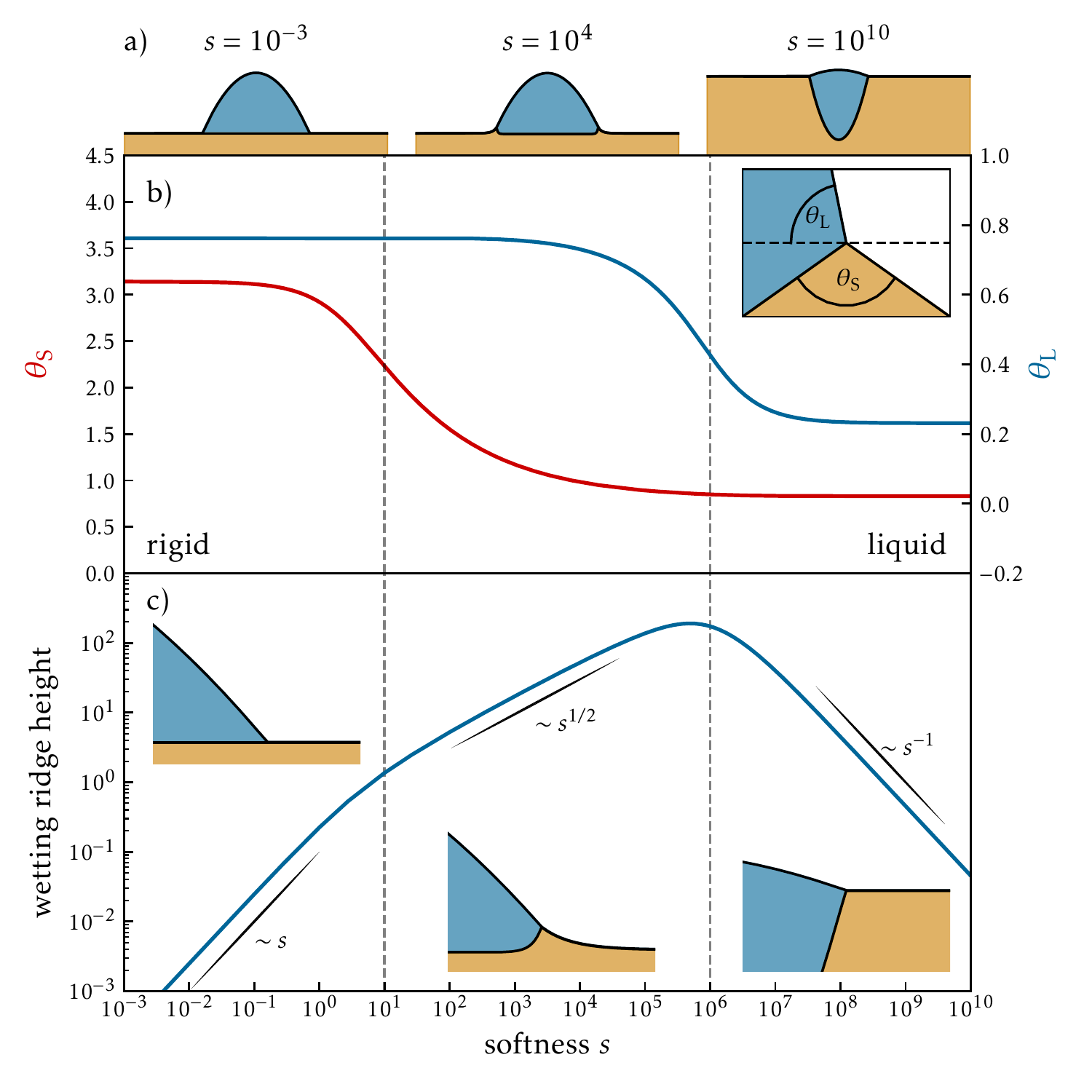}
	\caption{The sub-panels in (a) show the typical equilibrium drop and substrate morphologies (starting on the left) in the rigid limit $s=10^{-3}$, for intermediate softness $s=10^4$ and the liquid limit $s=10^{10}$. Panel (b) displays the dependence of the angles $\theta_\text{L}$  and $\theta_S$ at the wetting ridge (defined in the inset) on the substrate softness $s=\kappa^{-1}$ for a drop of volume $V \approx 10^6$ at $\sigma=0.1$. The two distinct transitions are discussed in detail in the main text. The vertical dashed lines indicate the values for the first two profiles in (a).  Panel (c) shows the dependence of the wetting ridge height on $s$ and indicates different scaling regimes. Note that the $\sqrt{s}$ scaling in the intermediate regime confirms the definition of the elastocapillary length \eqref{eq:lengths}. A corresponding video showing all steady states of the continuation in  $s$ can be seen in video~1 in the Supplementary Material.}
	\label{fig:DT_sketch}
\end{figure}

First, we investigate a static situation, namely, the dependence of the properties of sitting drops on the substrate softness $s=\kappa^{-1}$. As described in section~\ref{sec:num} this is done by employing numerical continuation to track solutions to the nondimensional steady equations~(\ref{eq:statics-nondim-h}) and~(\ref{eq:statics-nondim-xi}). Typical drop and substrate profiles are given in Fig.~\ref{fig:DT_sketch}~(a), showing how the equilibrium morphology changes as the substrate changes from very rigid to very soft. Here we explore the essential features, and compare the results to existing literature.

Of particular interest are the angles sketched in the inset of Fig.~\ref{fig:DT_sketch}~(b), namely, $\theta_\text{L}$, the angle the liquid-gas interface takes with the horizontal and $\theta_S$, the opening angle of the wetting ridge. They are expected to smoothly change from fulfilling Young's law for a rigid substrate ($s\to0$) to fulfilling Neumann's law in the limiting case of a liquid substrate ($s\to\infty$). The change of both angles with softness is given in Fig.~\ref{fig:DT_sketch}~(b). As in the full elastic theory for drops with contact angle $\pi/2$ in Ref.~\cite{LWBD2014jfm} two distinguished separate transitions occur. Starting in the rigid limit and increasing the softness, first, one observes the formation of wetting ridges. As the overall shape of the drop itself barely changes, in this first transition $\theta_\text{L}$ remains approximately constant while $\theta_S$ strongly decreases (in Fig.~\ref{fig:DT_sketch}~(b) at about $s=1\dots10^2$). In parallel, the wetting ridge grows until it approaches a maximal height, as shown in Fig.~\ref{fig:DT_sketch}~(c). Increasing the softness further, the substrate below the drop is compressed owing to the Laplace pressure imposed by the drop. In consequence, the drop gradually sinks into the elastic layer accompanied by an inwards rotation of the wedge structure in the three-phase contact region. This rotation marks the second transition (in Fig.~\ref{fig:DT_sketch}~(b) at about $s=10^5\dots10^7$): now $\theta_\text{L}$ strongly decreases while $\theta_S$ remains approximately constant. The second transition is also reflected as a maximum in the wetting ridge height, visible in Fig.~\ref{fig:DT_sketch}~(c). 

The two steps of the double transition can be related to the ratio of the elastocapillary length $\ell_\mathrm{ec}$ to the relevant microscopic and macroscopic length scales of the system \cite{LWBD2014jfm}. Elasticity becomes important when $\ell_\mathrm{ec}$ becomes comparable to the microscopic length scale, here $\ell_\mathrm{ec} \sim h_\mathrm{a}$. For our nondimensional quantities this implies the first transition occurs when $\tilde\ell_\mathrm{ec}=\sqrt{\sigma/\kappa}=\sqrt{\sigma s}\approx 1$, i.e., for $s\approx 1/\sigma$. For $\sigma=0.1$ we have $s\approx 10$ in agreement with Fig.~\ref{fig:DT_sketch}~(b).  Upon entering the  intermediate regime the ridge height $\sim s^{1/2}$, which implies that the ridge height is proportional to $\tilde \ell_{\rm ec}$. Likewise, the second transition emerges when the droplet sinks in, which happens when $\tilde \ell_\mathrm{ec}$ reaches to the macroscopic scale of the droplet size, here the radius. In nondimensional terms this occurs when $\tilde\ell_\mathrm{ec}\approx \sqrt{V}$ where $V$ is the nondimensional volume of the 2d droplet,\footnote{Note that $V$ is defined as the drop volume above the adsorption layer of nondimensional thickness one and is strictly speaking an area.}, i.e., $\sqrt{\sigma/\kappa}\approx \sqrt{V}$, i.e., $s \approx V/\sigma$. For $\sigma=0.1$ and drop volume $V=10^6$ we have $s\approx 10^7$. This is close to the value observed in Fig.~\ref{fig:DT_sketch}~(b).

\begin{figure}[tbh]
		\includegraphics[scale=1]{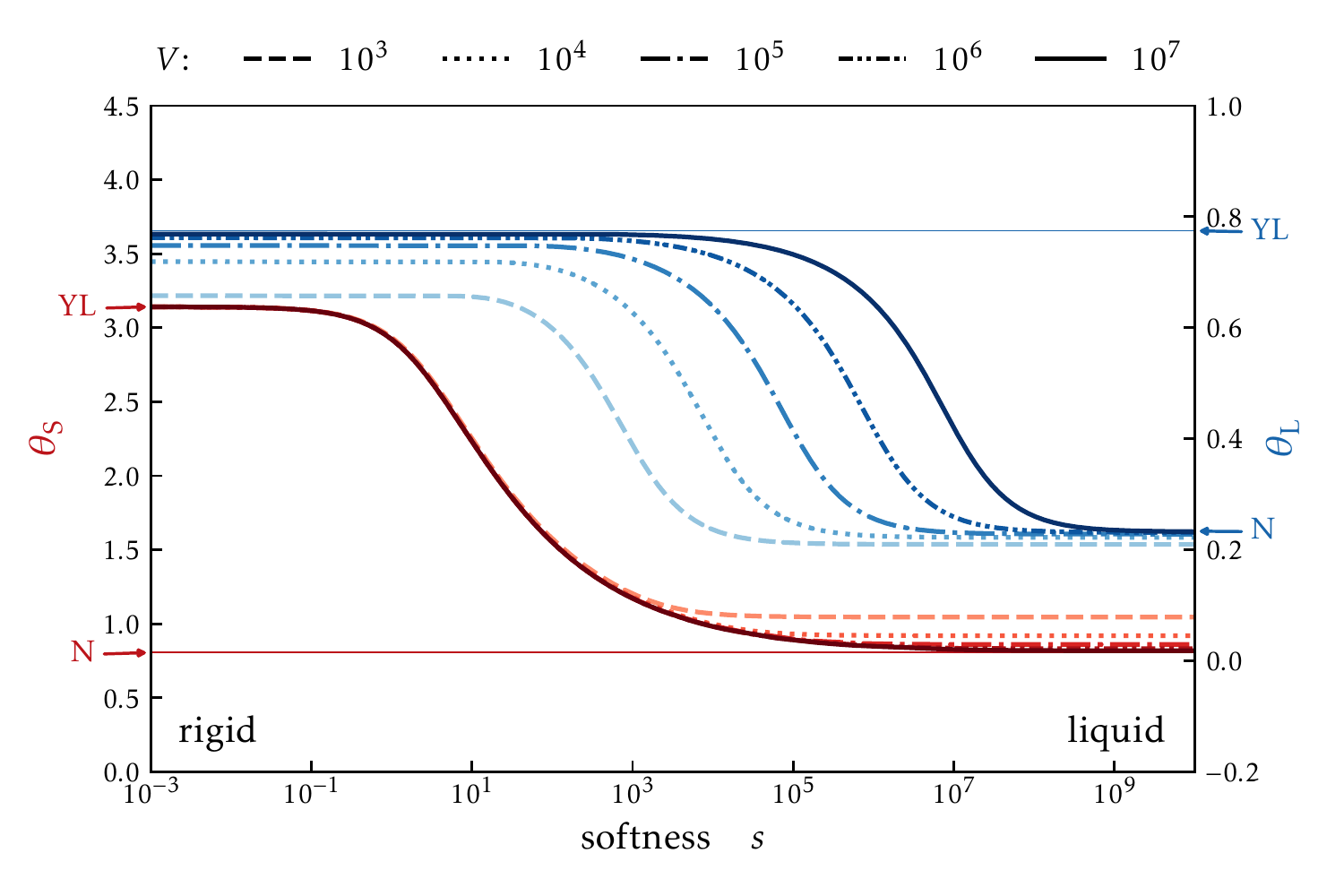}
                \caption{Shown is the dependence of the angles $\theta_\text{L}$ and $\theta_S$ [see inset of Fig.~\ref{fig:DT_sketch}~(b)] on substrate softness $s=\kappa^{-1}$ for steady drops of different volumes $V$ as given in the legend. The remaining parameters are as in Fig.~\ref{fig:DT_sketch}. As the volume increases, the two distinguished transitions become increasingly separated and for the angles the approach of the Young-Laplace (YL) and Neumann (N) law in the rigid and liquid limit, respectively, improves. The respective analytically predicted values are marked by arrows and thin horizontal lines.
                  }
	\label{fig:DT_volumes}
      \end{figure}
      
\begin{figure}[tbh]
		\includegraphics[scale=1]{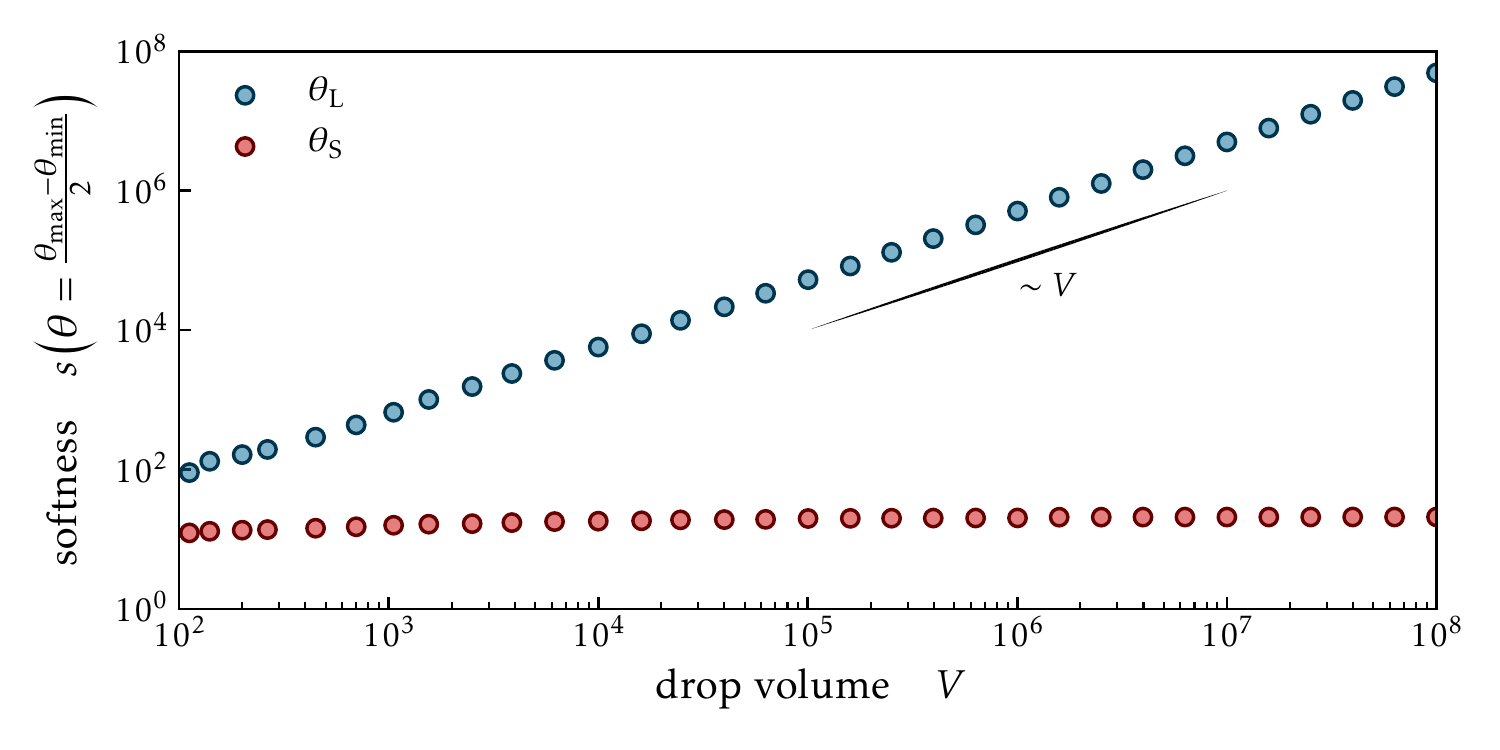}
	\caption{Softness values where the two distinguished transitions in the angles $\theta_\text{L}$ and $\theta_S$ in dependence of drop volume occur [cf.~Fig.~\ref{fig:DT_volumes}]. The transition values are determined as the values where the respecting angle is $\theta=(\theta_\text{max}-\theta_\text{min})/2$.}
	\label{fig:DT_shift}
\end{figure}

The dependence of the second transition on the drop volume is further scrutinised in Fig.~\ref{fig:DT_volumes} where the dependencies of the two angles $\theta_\text{L}$ and $\theta_S$ on softness $s$ are given for six different volumes. The location of the first transition at $s\approx10$ is clearly independent of drop volume. The location of the second transition, however, does depend on the droplet volume. To quantify this we show in Fig.~\ref{fig:DT_shift} the position of the two transitions (measured as $s$-value where $\theta=(\theta_\text{max}-\theta_\text{min})/2$) as a function of softness. Indeed, the transition value for $\theta_S$ remains nearly constant while for $\theta_\text{L}$ it increases proportional to drop volume. With the condition $\tilde\ell_\mathrm{ec}=\sqrt{\sigma s}\approx \sqrt{V}$ we indeed expect $s\sim V/\sigma$ in agreement with the full elastic theory \cite{LWBD2014jfm}.

Finally, we return to Fig.~\ref{fig:DT_volumes} and inspect the asymptotic values of the contact angles in the rigid ($s\to0$) and liquid ($s\to\infty$) limits. One notices that for large volumes the limiting angles are independent of droplet size, and these values perfectly agree with the expected values from Young's law and Neumann's law as indicated by the arrows in Fig.~\ref{fig:DT_volumes}.
With decreasing volume, however, we observe an increasing deviation from these ``macroscopic'' values which reflects the mesoscopic character of the model, i.e., the modelling of wettability using a wetting potential. However, macroscopic results of elastic wetting are correctly captured when drops are much larger than the thickness of the  adsorption layer.

%%%%%%%%%%%%%%%%%%%%%%%%%%%%%%%%%%%%%%%%%%%%%%%%%%%%%%%%%%%%%%%%%%%%%%%%%%%%%%%
\section{Drop spreading} \label{sec:spreading}
%%%%%%%%%%%%%%%%%%%%%%%%%%%%%%%%%%%%%%%%%%%%%%%%%%%%%%%%%%%%%%%%%%%%%%%%%%%%%%%

We now investigate the dynamical process of droplet spreading. Of particular interest is the effect of viscoelastic braking, i.e., the decrease of contact line speed during spreading with increasing substrate softness \cite{CaSh2001l,LoAL1996lb,KDGP2015nc,ChKu2020sm,AnSn2020arfm}. Any motion of the contact line will displace the wetting ridge, which induces a time-dependent deformation. Owing to the viscoelasticity of the solid, this offers an additional source of dissipation that slows down the spreading of the drop. Here we verify whether this viscoelastic braking is correctly accounted for in our model. Beyond that, we compare the case of a drop of partially wetting liquid, i.e., employing the wetting energy (\ref{eq:energy-wet-f}) as in section~\ref{sec:steady}, and the case of completely wetting liquid. The latter is realized by using a positive sign for the long-range contribution to the wetting energy (\ref{eq:energy-wet-f}) resulting in a strictly repulsive disjoining pressure. 
As described in section~\ref{sec:num} the spreading is analysed by direct time simulation of Eqs.~(\ref{eq:dynamics-nondim-h}) and~(\ref{eq:dynamics-nondim-xi}). Note that here we make use of the reflection symmetry w.r.t.\ the maximum of the drop, and only simulate one half of the drop using Neumann boundary conditions.

\begin{figure}
	\includegraphics[scale=1]{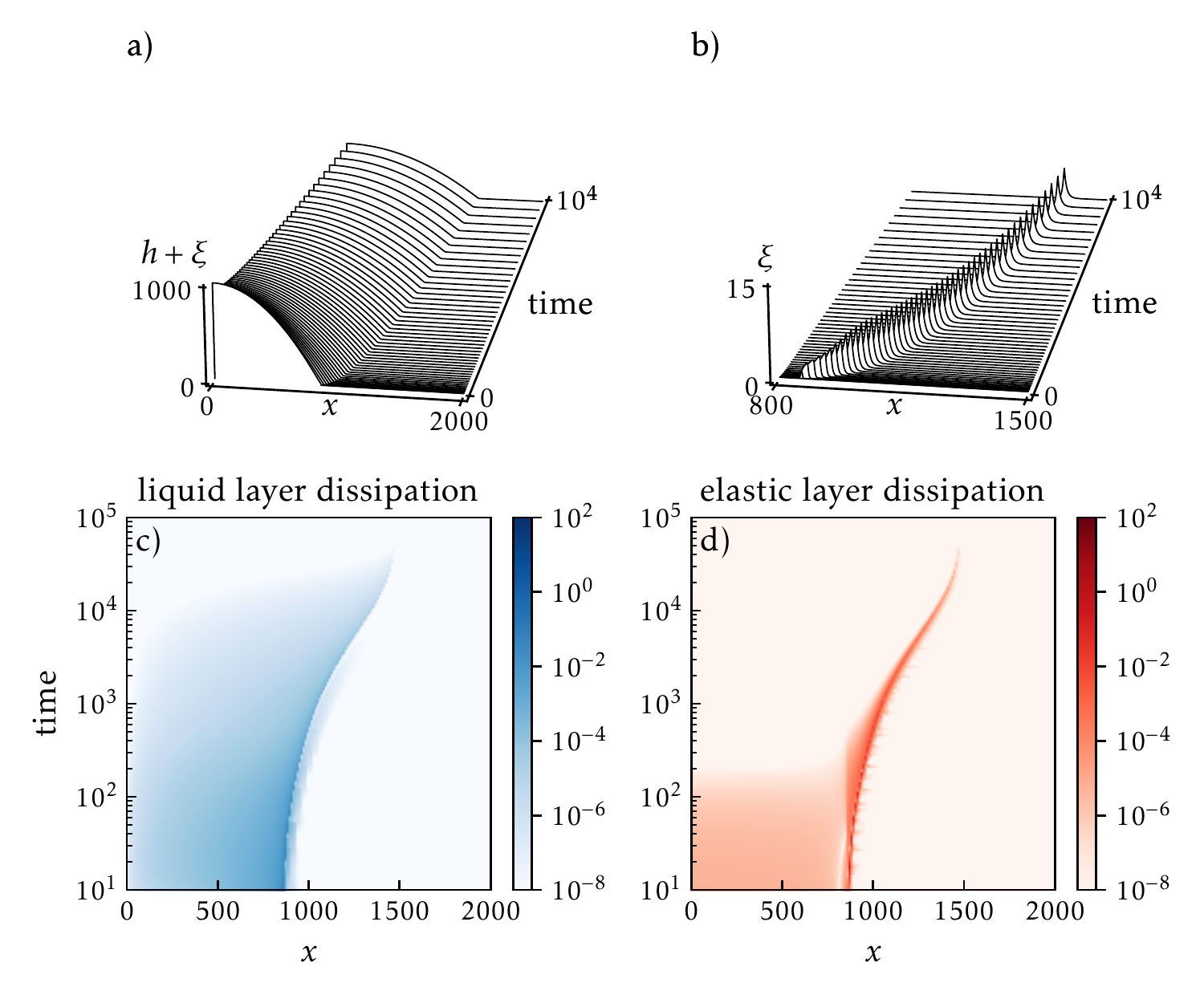}
	\caption{A typical time evolution of a spreading drops of partially wetting liquid at intermediate softness $s=10^2$ is characterized by space-time plots of the (a) drop and (b) wetting ridge profiles, and by the spatially resolved dissipation in (c) the liquid and (d) the substrate. The remaining parameters are $D=2000$, $V\approx10^6$, $\tau=1$, and $\sigma=0.1$. A corresponding time evolution can be seen in video~2 of the Supplementary Material.}
\label{fig:spreading-single}
\end{figure}

 \subsection{Partial wetting}
 
 The initial profile is a drop of parabolic shape on a flat substrate with an initial contact angle that is three times larger than the equilibrium contact angle resulting from Young's law, i.e., $\tilde{\theta}_\mathrm{ini}=3\tilde{\theta}_\mathrm{eq} = 3\sqrt{3/5}$. The initial drop height is $h_\mathrm{max}=1000$, i.e., three orders of magnitude larger than the adsorption layer height. A typical time evolution at intermediate softness $s=10^2$ is shown in Fig.~\ref{fig:spreading-single}. Given that the initial contact angle is larger than its equilibrium value, the disjoining pressure will drive a flow of liquid towards the contact line. This induces a contact line motion towards equilibrium, as visible in Fig.~\ref{fig:spreading-single}~(a). The corresponding evolution of the wetting ridge is indicated in Fig.~\ref{fig:spreading-single}~(b). At very early time, a ridge is growing out of the initially flat substrate and subsequently moves along with the contact line. 
 This behavior closely resembles that observed in experiments \cite{GKAS2020sm}. The corresponding dissipations in the liquid and in the substrate are reported, respectively, in Fig.~\ref{fig:spreading-single}~(c) and (d). For both, the largest dissipation happens near the contact line. For this set of parameters, it turns out that the dissipation peak is larger in magnitude inside the solid, showing that the substrate dissipation dominates over liquid dissipation. 
 This is the viscoelastic braking effect that we quantify in detail below.

A common way to characterize dynamical wetting is via the dependence of the (dynamic) liquid angle $\theta_L$ on contact line velocity $v$. To extract this information from our simulations we trace the estimated macroscopic contact line position $x_C$, and its velocity $v=dx_C/dt$, during the spreading process. There are different ways of defining $x_C$ in the mesoscopic model.%
\footnote{Common approximations for $x_C$ are: (i) the use of the position of steepest slope of the liquid-gas interface ($\theta_\text{L}$ is obtained as slope of the corresponding tangent), (ii) fitting a parabola to the drop apex and evaluate $x_C$ and $\theta_\text{L}$ at its intersection with the reference flat liquid-elastic interface at $z=0$, (iii) the use of the intersection of the tangent from (i) and $z=0$, and (iv) position of the maximum of the wetting ridge. Our tests show that method~(i) appears to often shift $x_C$ into the droplet by a non-negligible amount. Further, we find method (ii) to be unreliable as it entirely ignores the local geometry of the three-phase contact region, a problem partly shared by method~(iii). Finally, method~(iv) is problematic when the contact line surfs on the wetting ridge, i.e.\ when the wetting ridge falls behind. Beyond that, method~(iv) requires a fitting routine to prevent errors resulting from the discretization. Ref.~\cite{ChKu2020sm} uses method~(iii) to predict certain subtle differences in the $x_C(t)$-dependence of the spreading of partially and completely wetting liquids on softness. Here, we are only considering the dependence of dynamic contact angle on contact line speed $dx_C/dt$ and method~(iii) gives robust results.}
Here, we determine $x_C$ as the intersection point of the tangent to the liquid-gas interface at the position of its steepest slope with the reference flat liquid-elastic interface at $z=0$.

\begin{figure}[tbh]
\includegraphics[width=0.9\hsize]{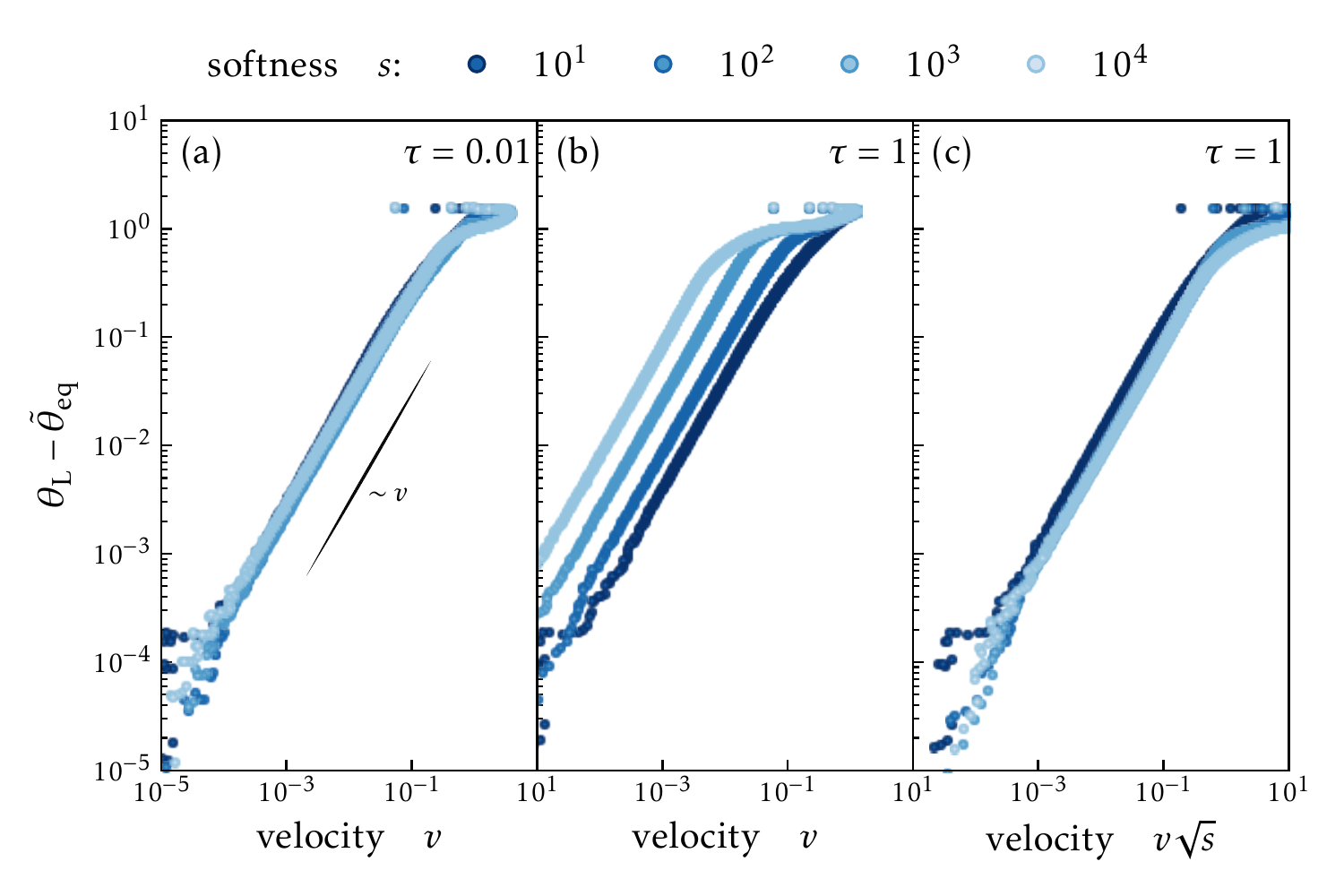}
\caption{Dynamic contact angles, characterized as $\Delta\theta_\text{L} = \theta_\text{L}-\tilde\theta_\mathrm{eq}$, over contact line velocity $v=dx_C/dt$. (a) Result for fast relaxation of the substrate ($\tau=0.01$), for which the spreading dynamics is hardly affected when varying the softness $s$. The spreading dynamics is governed by dissipation in the liquid. (b) Result for slow relaxation of the substrate ($\tau=1$), for which the spreading dynamics is strongly affected when varying the softness $s$. (c) Same data as in (b), but using the rescaled velocity $v\sqrt{s}$ (see text for motivation). The collapse of data shows that the dynamic angle is entirely governed by the solid dissipation, which is the hallmark of viscoelastic braking. The indicated power laws are discussed in the main text.}
\label{fig:spreading-partial}
\end{figure}

Figure ~\ref{fig:spreading-partial} presents the resulting dynamic liquid contact angles, for different softness (varying $s$) and different degrees of substrate viscoelasticity (varying $\tau$). Specifically, we report the difference in angle with respect to its equilibrium value,  $\Delta\theta_\text{L} = \theta_\text{L}-\tilde\theta_\mathrm{eq}$, as a function of the contact line velocity $v=d x_C/d t$.
In all cases there is an extended regime of proportionality $\Delta\theta_\text{L} \sim v$ where the change in contact angle is linear with the contact line velocity. Such a linear dependence is indeed expected here. The liquid dynamics is governed by Stokes flow, for which the dynamic contact angle in partial wetting is linear with velocity \cite{Duss1979arfm,TVNB2001pre,BEIM2009rmp}. Similarly, a Kelvin-Voigt solid (as in the present case), will give a linear dependence of the dynamic contact angle \cite{KDGP2015nc}. We remark that in experiments, the exponent is much lower than unity, with typical values around 0.5. This is because the dynamic contact angle inherits the scaling of the loss modulus -- for the Kelvin-Voigt model this is linear in frequency but in crosslinked polymer networks typically exhibit smaller exponents \cite{AnSn2020arfm}. 

However, we still need to disentangle the relative roles of liquid and solid dissipation. Figure~\ref{fig:spreading-partial}~(a) corresponds to a very small value of $\tau$, so that the solid very quickly adapts to any changes in the liquid. We observe that the spreading dynamics is completely independent of the softness $s$ in this case. This implies that the substrate has no influence on the spreading, in spite of the presence of a wetting ridge. Indeed, at very small $\tau$ the solid's response is nearly instantaneous, which corresponds to a perfectly elastic limit where there is no dissipation inside the substrate. By contrast, for larger values of $\tau$ the softness plays an important role, as seen in Fig.~\ref{fig:spreading-partial}~(b). Softer substrates (larger $s$) give a larger departure from the equilibrium angle. This indeed is a signature of the viscoelastic braking effect: for a given value of $\theta_L-\theta_\mathrm{eq}$, the contact line velocity is lower on softer substrates. To verify whether the substrate dissipation indeed dominates over liquid dissipation, we try to collapse the same data using a characteristic velocity of the substrate. Following \cite{KDGP2015nc,AnSn2020arfm}, this velocity is given by the ratio $\ell_\mathrm{ec}/\tau_\xi$, which in the present scaling implies a dependence on softness as $\sim 1/\sqrt{s}$. Figure ~\ref{fig:spreading-partial}~(c) therefore shows the dynamic contact angle plotted against the rescaled dimensionless substrate velocity $v\sqrt{s}$: the data now collapses, confirming that the spreading dynamics is indeed completely governed by the substrate's viscoelasticity.

 \subsection{Complete wetting}

\begin{figure}
	\includegraphics[scale=1]{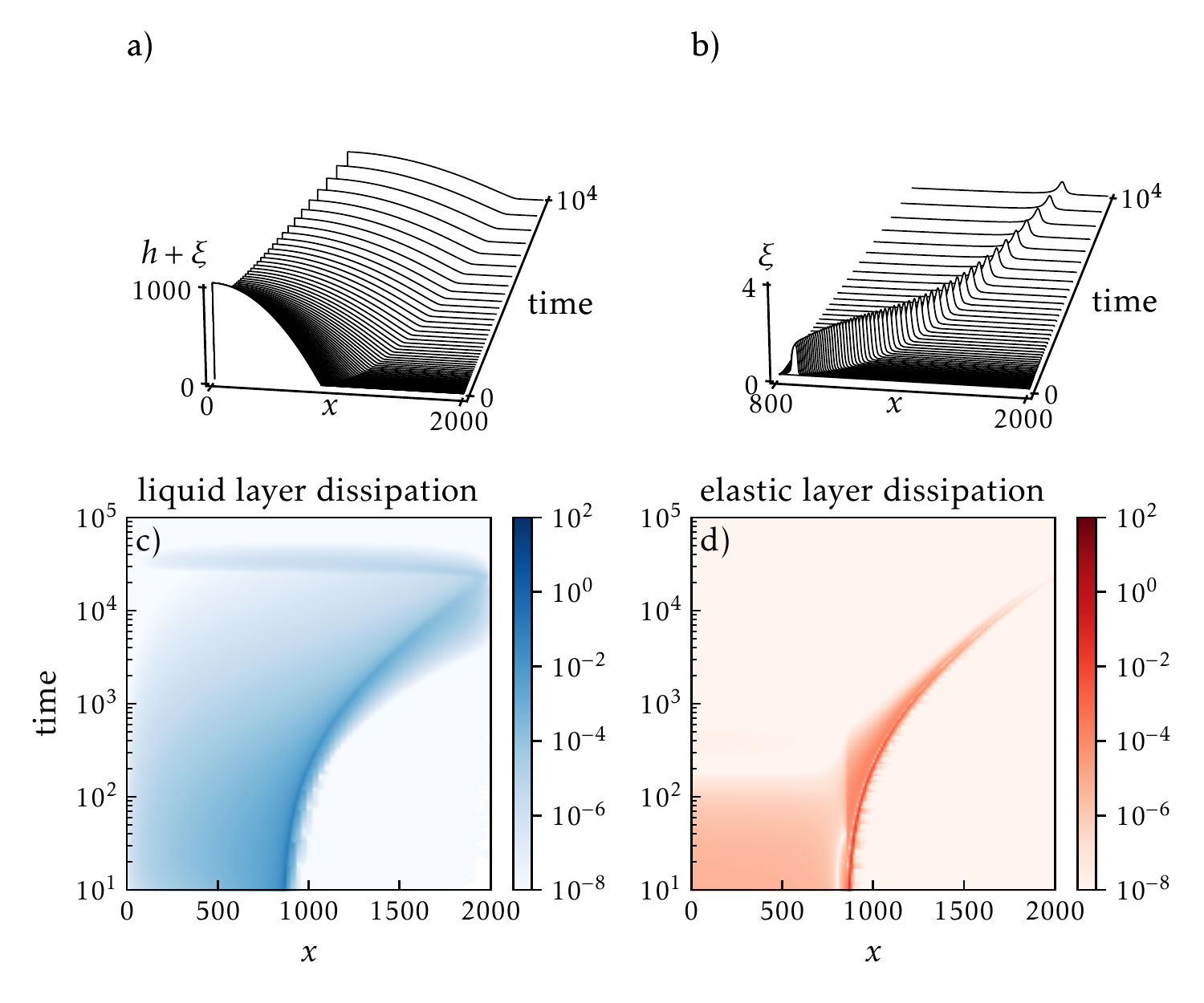}
	\caption{A typical time evolution of a spreading drops of completely wetting liquid at intermediate softness $s=10^2$ is characterized by space-time plots of the (a) drop and (b) wetting ridge profiles, and by the spatially resolved dissipation in (c) the liquid and (d) the substrate. The remaining parameters are $D=2000$, $V\approx10^6$, $\tau=1$, and $\sigma=0.1$.  A corresponding time evolution can be seen in video~3 of the Supplementary Material.
          	}
	\label{fig:spreading-single-wetting}
\end{figure}

The case of a completely wetting fluid can be analogously treated, by using a positive instead of the negative sign in the wetting energy~(\ref{eq:energy-wet-f}). A corresponding typical time evolution is displayed in Fig.~\ref{fig:spreading-single-wetting}, for the same initial conditions as for the partially wetting case. At early times, the dynamics is very similar to that of the partially wetting case. The liquid starts to spread and a ridge is pulled out of the initially flat solid, however, the ridge remains quite small. Owing to the complete wetting, however, the drop does not spread toward a shape with finite equilibrium angle but, instead, the contact line continues to spread. During this process the wetting ridge gradually decays, cf.~Fig.~\ref{fig:spreading-single-wetting}~(b), in contrast to the growth observed under partial wetting conditions. We attribute this trend to the continuous decrease of the contact angles reached during spreading, which diminishes the upward pulling force of surface tension. In consequence, the dissipation in the substrate decays more rapidly than the dissipation in the fluid, as seen in Fig.~\ref{fig:spreading-single-wetting}~(c,d).

\begin{figure}[tbh]
		\includegraphics[scale=1]{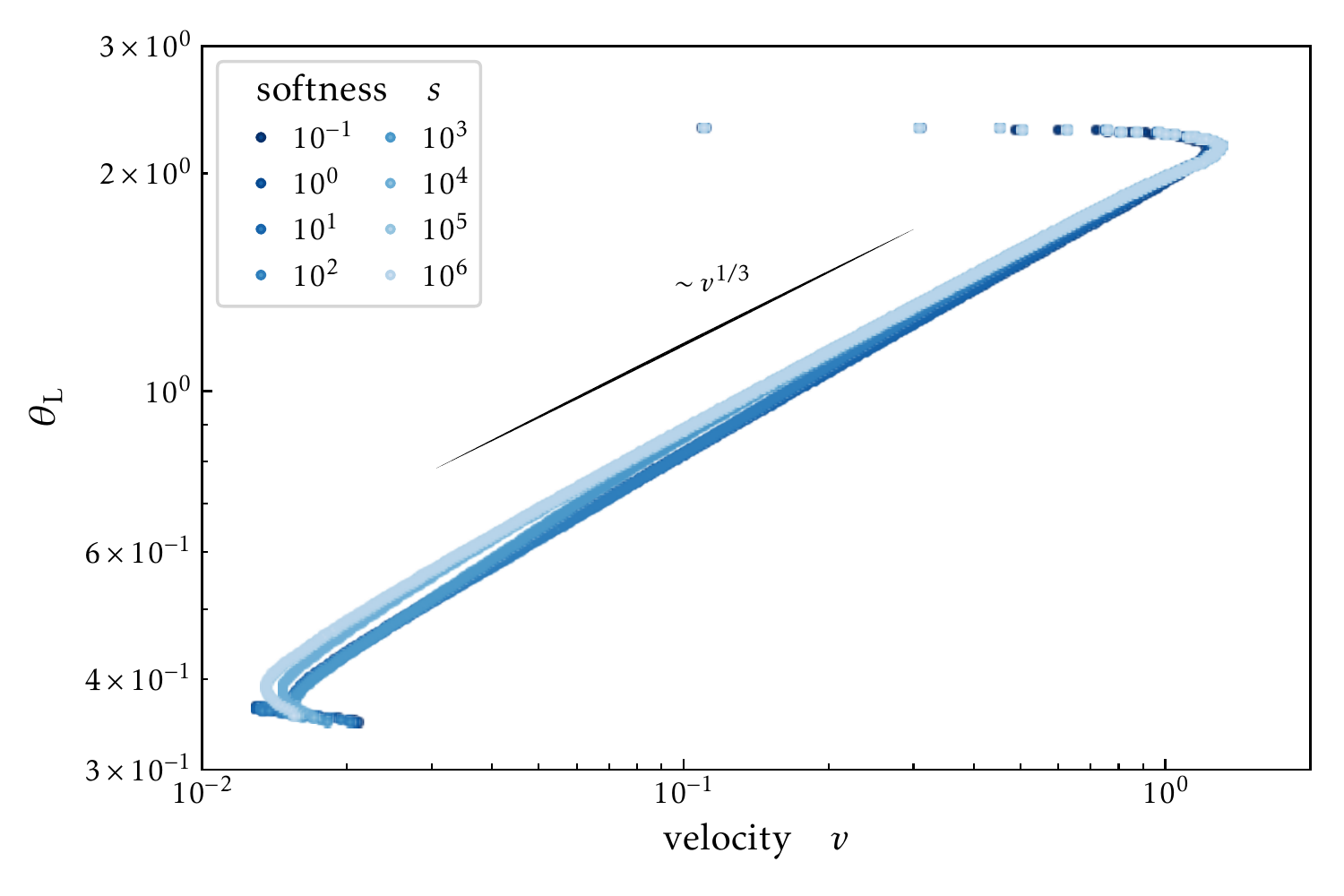}
                \caption{For the completely wetting case the spreading dynamics is characterised by the dependency of the difference $\Delta\theta_\text{L} = \theta_\text{L}-\tilde\theta_\mathrm{eq}$ of the dynamic contact angle and its equilibrium value on the contact line velocity $v=d x_C/d t$. Results are given for different softness $s=\kappa^{-1}$ as given in the legend. The remaining parameters are as in Fig.~\ref{fig:spreading-single}. 
}
\label{fig:spreading-wetting}
\end{figure}

Figure~\ref{fig:spreading-wetting} shows the dynamic contact angle $\theta_\mathrm{L}$ against the spreading velocity $v$ for various softness $s$. The chosen value of the relaxation $\tau=1$ exhibits a strong viscoelastic braking in the partially wetting conditions -- the complete wetting case, by contrast, shows no significant dependence on the substrate softness $s$. This reflects the observations in Fig.~\ref{fig:spreading-single-wetting}, where only a comparatively small and soon decaying wetting ridge is found. The observed relation for the dynamic angle is consistent with the scaling $\theta_L \sim v^{1/3}$, which is the classical Tanner-Cox-Voinov relation \cite{Tann1979jpd,BEIM2009rmp} for viscous spreading under complete wetting conditions. We thus conclude that in complete wetting, the softness of the substrate does not influence the spreading dynamics.

\clearpage      

%%%%%%%%%%%%%%%%%%%%%%%%%%%%%%%%%%%%%%%%%%%%%%%%%%%%%%%%%%%%%%%%%%%%%%%%%%%%%%%
\section{Droplet coarsening}\label{sec:coarse}
%%%%%%%%%%%%%%%%%%%%%%%%%%%%%%%%%%%%%%%%%%%%%%%%%%%%%%%%%%%%%%%%%%%%%%%%%%%%%%%

When an ensemble of drops coarsens the average drop size and distance increase in time while their number decreases. This may occur at overall fixed liquid volume \cite{LiGr2003l} or involve condensation (and evaporation) \cite{FrKB1991pra}. In any case, the underlying elementary process is the coarsening of two (and sometimes three) drops into one. For such a coarsening step, there exist two basic coarsening modes.

This is, on the one hand, the mass transfer mode of coarsening: transport of material occurs from one droplet to the other one(s) while the centers of mass of the drops do not move. The mass flux is always directed from the small drop to the larger one (guaranteeing coarsening) due to the difference in Laplace pressure in the two coarsening drops. The process ends when the smaller drop has vanished and all the mass is contained in the remaining drop. There are different transport channels for the mass transfer mode: transport may occur via the vapour phase (for volatile liquids) or through an ultrathin adsorption layer (for nonvolatile, partially wetting liquids). This coarsening mode is in the literature referred to as volume or mass transfer mode, (drop) collapse mode, diffusion-controlled ripening, or Ostwald ripening \cite{PiPo2004pf,GlWi2005pd,GORS2009ejam,Dai2010n}.

On the other hand, there is the translation mode of coarsening: the entire drops  migrate toward each other, until their contact lines touch and the drops fast coalesce. This mode is also referred to as collision, coalescence, or migration mode \cite{PiPo2004pf,GlWi2005pd,GORS2009ejam,Dai2010n}.
Sketches of mass transfer and translation modes can be found in~\cite{GORS2009ejam} and their relation to the translation symmetry mode for single fronts or contact lines in a homogeneous system is discussed in section~3.4 of \cite{Thie2007chapter}. The stabilization of both coarsening modes by substrate heterogeneities gives rise to intricate bifurcation behaviour \cite{TBBB2003epje}.

The two coarsening modes can contribute to the coarsening process in a mixed way \cite{PiPo2004pf,GlWi2005pd} and can nearly not be distinguished in the resulting scaling laws describing the increase of mean drop volume and drop distance for coarsening large drop ensembles. For the most common cubic mobility function (resulting from no-slip boundary conditions at the solid substrate \cite{OrDB1997rmp,MuWW2005jem}) the scaling laws for the two modes only differ by a logarithmic factor \cite{GORS2009ejam}. Reference~\cite{GlWi2005pd} shows for drops on a one-dimensional solid substrate that beyond a certain threshold with increasing mean drop size and decreasing mean drop distance the translation mode becomes more prevalent until it finally dominates (cf.~Fig.~15 of \cite{GlWi2005pd} and also \cite{GlWi2003pre,Glas2008sjam}). Below the threshold, only the mass transfer mode is found.

Here we briefly show that the developed model is well suited to study drop coarsening on soft substrates. In particular, the following two sections discuss the influence of softness on the individual coarsening process of two drops and on the collective coarsening behaviour of a large ensemble, respectively.

%%%%%%%%%%%%%%%%%%%%%%%%%%%%%%%%%%%%%%%%%%%%%%%%%%%%%%%%%%%%%%%%%%%%%%%%%%%%%%%
\subsection{Case of two drops} \label{sec:coarse2}
%%%%%%%%%%%%%%%%%%%%%%%%%%%%%%%%%%%%%%%%%%%%%%%%%%%%%%%%%%%%%%%%%%%%%%%%%%%%%%%
%
\begin{figure}
	\includegraphics[width=0.8\hsize]{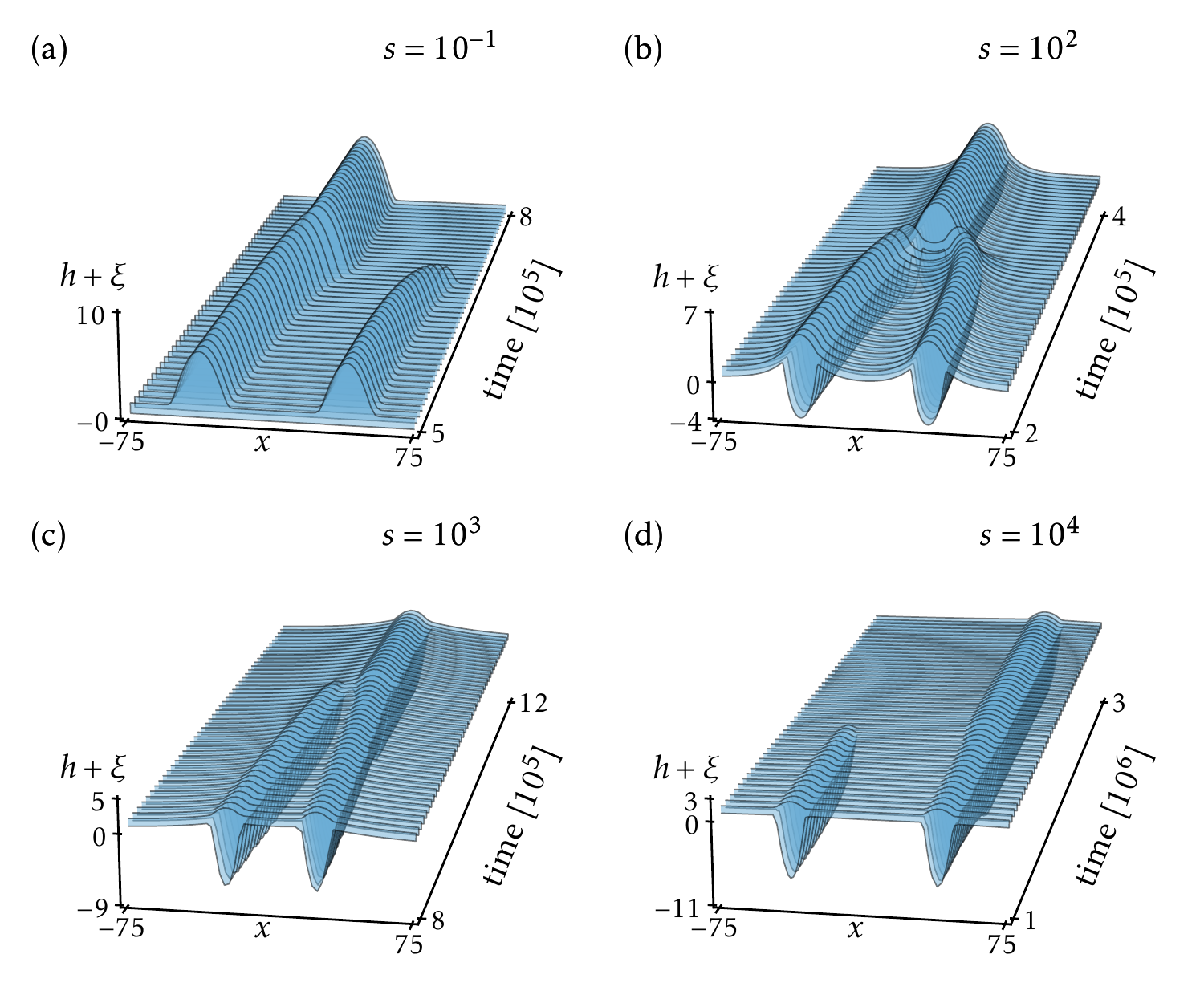}
	\caption{Shown are space-time plots for the coarsening of two drops of partially wetting liquid into a single drop. The softness increases from (a) $s=10^{-1}$ in the rigid limit, via (b) $s=10^2$ and (c) $s=10^3$ (intermediate elasticity) to (d) $s=10^4$ (liquid limit). With increasing softness the dominant coarsening mode changes as discussed in the main text. 
		The volume of each drop is $V=100$, the initial distance of their ``inner'' contact lines is $L_0=50$ and the domain size is $D=400$. The remaining parameters are as in Fig.~\ref{fig:spreading-single}. The corresponding time evolutions can be seen in videos~4 to~7 of the Supplementary Material, respectively.}
	\label{fig:twodrop-profiles}
\end{figure}

      Here, we employ the developed model to investigate the dynamics of the individual coarsening step for two neighboring drops and discuss how the dominant coarsening mode depends on substrate softness. The process is initiated with two parabolic drops of identical volume $V$ that are placed at a distance $L_0$ (between the contact lines) on an initially flat but soft substrate. Their contact angle corresponds to the equilibrium value in the corresponding rigid limit.  The resulting profile is perturbed by some small initial noise to slightly break the initial reflection symmetry about the center line ($x=0$). Typical time evolutions for selected softnesses $s$ at fixed $V=100$ and $L_0=50$ are given in Fig.~\ref{fig:twodrop-profiles} as space-time plots. In all cases, initially, up to about $t=O(10^2)$ the substrate-liquid interface relaxes, i.e., the drops sink into the soft substrate (not shown). The amount of sinking, i.e.\ the size of the indentation in the substrate, and the size and height of the developing wetting ridge all depend on softness (cf.~Section~\ref{sec:steady}). Then, the coarsening process sets in on a much larger time scale. Depending on softness, it is usually terminated at $t=10^5\dots10^7$. \

Inspecting Fig.~\ref{fig:twodrop-profiles}, we find that the coarsening process qualitatively changes with softness in a nonmonotonic manner. In panels (a) and (d), respectively corresponding to the rigid and soft limits, the droplets exchange volume with minor or without lateral translation (volume mode). By contrast, the intermediate softness of panels (b) and (c) exhibit a clear lateral motion of the drops towards one another (translation mode). This nonmonotonic coarsening is reminiscent of the nonmonotonic change of the wetting ridge height with softness, observed for static drops in Fig.~\ref{fig:DT_sketch}~(c). This suggests that the translation regime is mediated by elastic deformation, as a form of the Cheerios effect \cite{KPLW2016pnasusa}.

\begin{figure}
		\includegraphics[width=0.9\hsize]{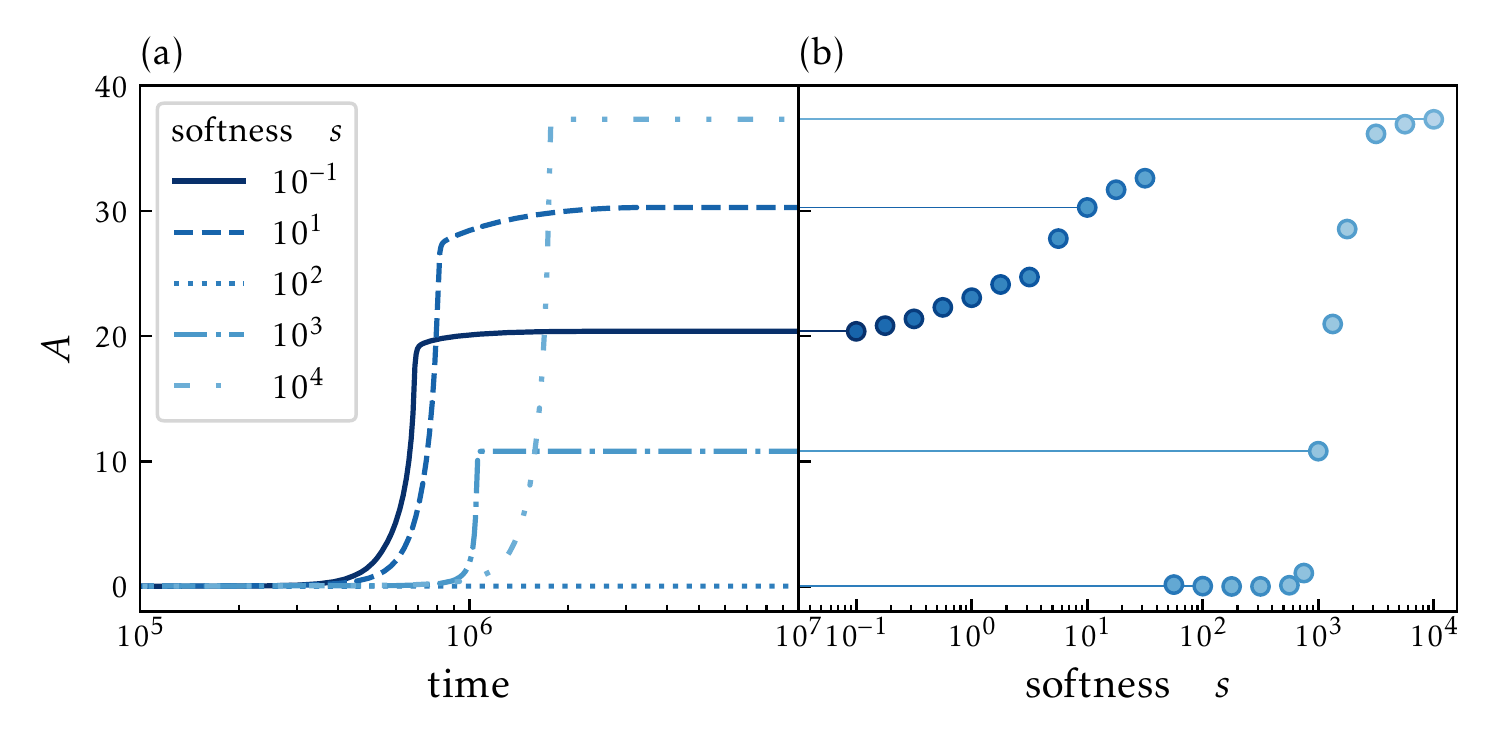}
  \caption{(a) Shown is the temporal change in the asymmetry measure $A(t)$ [Eq.~(\ref{eq:asym})] for two-drop coarsening processes at softness values given in the legend (including the values of Fig.~\ref{fig:twodrop-profiles}). The maximal possible value for the given drop volume and distance corresponds to the absolute value of the initial drop position $A_\mathrm{max}=|x_\mathrm{init}|\approx40$. Panel~(b) gives the intricate dependence of the final value $A_\mathrm{fin}$ on softness $s$. It shows that the volume mode is dominant apart from a finite range of intermediate elasticity ($s\approx10^2\dots10^3$) where the translation mode dominates. 
  See main text for further discussion.}
\label{fig:twodrop-asym}
      \end{figure}

To quantify the relative importance of  volume and translation mode, we introduce an asymmetry measure for the considered two-drop configuration, namely,
      \begin{equation}
        A(t)= \frac{1}{2V}\int_D x [h(x,t)-h_a]\, dx.
        \label{eq:asym}
      \end{equation}
      which corresponds to the center of mass of the drop volume above the adsorption layer ($h_a=1$). A coarsening process exclusively involving the translation mode that symmetrically moves the drops towards each other would keep $A=0$ as the center of mass remains at $x=0$. In contrast, an exclusive transfer of liquid from one to the other drop via the volume transfer mode would result in a maximal increase to $A=A_\mathrm{max}$ as all the volume is finally centered about the initial position of the remaining drop. If both coarsening modes contribute the resulting value of $A$ lies between the two extremes.

For the time evolutions in Fig.~\ref{fig:twodrop-profiles}, the corresponding measures $A(t)$ are given in Fig.~\ref{fig:twodrop-asym}~(a) allowing us to quantitatively discuss the changes of the coarsening mechanism with increasing softness: For (nearly) rigid substrates [$s=10^{-1}$, Fig.~\ref{fig:twodrop-profiles}~(a)] the volume transfer mode dominates, as one drop shrinks and vanishes while the other one grows. The asymmetry measure $A(t)$ monotonically increases from zero and reaches a final value of $A_\mathrm{fin}\approx 22<A_\mathrm{max}\approx40$ (Fig.~\ref{fig:twodrop-asym})~(a). It is interesting to note that in parallel to the volume transfer both drops slightly migrate into the same direction, pointing from the growing to the shrinking drop. Even after the small drop vanishes the remaining drop briefly keeps traveling in that direction. This effect is well known for drops on rigid substrates, see e.g.\ Refs.~\cite{PiPo2004pf,GlWi2005pd}, and is related to liquid motion in the adsorption layer. 

     Increasing the softness to  $s=10$, volume transfer remains dominant. The amount of migration into the same direction decreases resulting in an increase of $A_\mathrm{fin}$ to about $30$. The increase can be well appreciated in Fig.~\ref{fig:twodrop-asym}~(b) where the final asymmetry measure $A_\mathrm{fin}$ is shown as a function of softness $s$.
Reaching the regime of intermediate elasticity at about $s=10^2$ [Fig.~\ref{fig:twodrop-profiles}~(b)] the behaviour abruptly changes and coarsening becomes entirely dominated by the translation mode with $A_\mathrm{fin}\approx0$, as seen from the dotted line in Fig.~\ref{fig:twodrop-asym}~(a). When further increasing $s$ by about one magnitude, the volume mode starts to become important again: Here, translation is still dominant during the early stages of the process, but volume transfer becomes relevant in the final phase [see $s=10^{3}$, Fig.~\ref{fig:twodrop-profiles}~(c), $A_\mathrm{fin}\approx 10$]. Then, in approaching the liquid limit at larger $s$, the translation mode loses all its impact and the volume mode again dominates coarsening. In contrast to the rigid limit, now no drop migration occurs [$s=10^{4}$, Fig.~\ref{fig:twodrop-profiles}~(d)], and $A_\mathrm{fin}$ approaches $A_\mathrm{max}$. The described intricate change from volume transfer to translation mode and back can best be appreciated in Fig.~\ref{fig:twodrop-asym}~(b). Independently, of the non-monotonic change in mode dominance, Fig.~\ref{fig:twodrop-asym}~(a) clearly indicates coarsening slows down with increasing softness.

\begin{figure}
	\includegraphics[width=0.9\hsize]{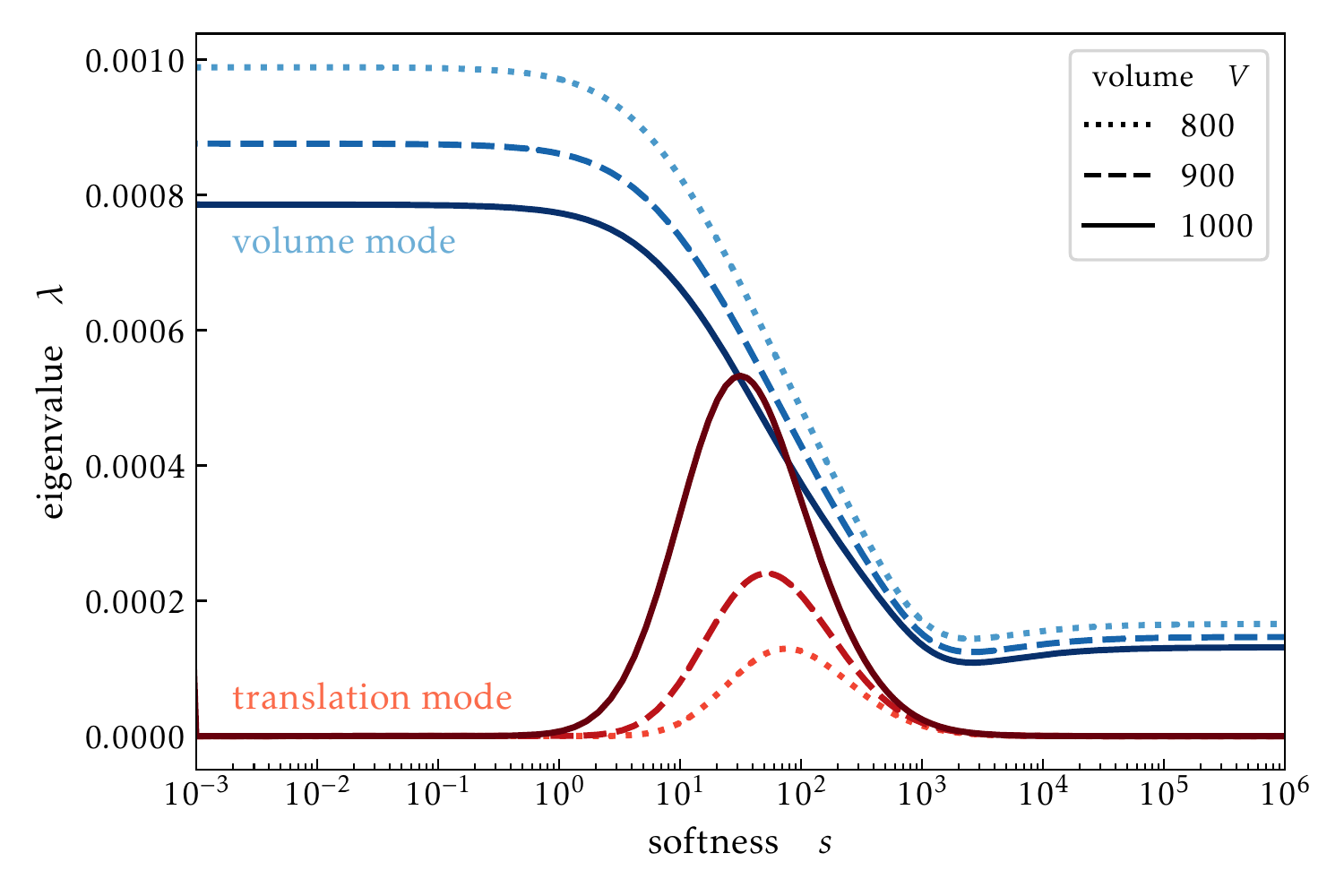}
	\caption{Eigenvalues $\lambda$ of the volume mode (blue) and the translation mode (red) for a periodic system of equally separated steady drops. Shown are results for the three different drop volumes $V$ as given in the legend. The domain size is fixed to $D=200$, i.e., a larger drop volume corresponds to a smaller drop distance. The volume mode weakens with increasing softness and drop size while the translation mode shows an increasingly pronounced maximum at intermediate elasticity that even exceeds the translation mode for the largest drop. The remaining parameters are as in Fig.~\ref{fig:spreading-single}.}
	\label{fig:twodrop-eigenvalues-softness}
\end{figure}

To gain a deeper understanding of the transition between modes, we perform a linear stability analysis for the alternative constellation of a periodic drop array. Namely, two identical drops of volume $V$ are placed in a domain of finite size $D=200$ at respective positions $x=0$ and $x=\pm D/2$. The corresponding unstable periodic steady two-drop state and its linear stability are continued (using \texttt{PDE2path}). Starting  at $s=10^{-3}$ in the rigid limit, the softness is increased up to $s=10^6$ in the liquid limit. The resulting real eigenvalues corresponding to volume transfer mode (the changes in film height profile are antisymmetric w.r.t.\ the center line between drops) and translation mode (symmetric changes in film height profile) of coarsening are presented in Fig.~\ref{fig:twodrop-eigenvalues-softness} for three selected $V$. The blue curves indicate the volume transfer mode, while red curves correspond to the translation mode. The volume mode is strongest in the rigid limit and decreases nearly monotonically with increasing softness (up to the small dip after the strong step-like decrease at intermediate softness $s=10^1\dots10^2$). It further slightly decreases overall with increasing drop volume, a change which is more pronounced in the rigid limit.

In contrast, the translation mode is nearly absent apart from intermediate softness where it reaches a maximum. This peak becomes more pronounced with increasing volume. In consequence, the translation mode eventually dominates the volume transfer mode for sufficiently large drops at intermediate $s$ in accordance with our time simulations Fig.~\ref{fig:twodrop-asym}~(b). 
The dominance of the translation mode in the range of intermediate softness approximately coincides with the intermediate regime for the ``double transition" of static drops discussed in section~\ref{sec:steady}, compare in particular, Fig.~\ref{fig:DT_shift} at $V=100\dots1000$ with Fig.~\ref{fig:twodrop-asym}~(b) and Fig.~\ref{fig:twodrop-eigenvalues-softness}. It is in this regime that a prominent wetting ridge occurs (cf.~Fig.~\ref{fig:DT_sketch}~(c)), through which the droplets interact via substrate deformations. Hence, we indeed conclude that the dominant translation mode is a direct consequence of the Cheerios effect.

%%%%%%%%%%%%%%%%%%%%%%%%%%%%%%%%%%%%%%%%%%%%%%%%%%%%%%%%%%%%%%%%%%%%%%%%%%%%%%%
\subsection{Case of drop arrays} \label{sec:coarseN}
%%%%%%%%%%%%%%%%%%%%%%%%%%%%%%%%%%%%%%%%%%%%%%%%%%%%%%%%%%%%%%%%%%%%%%%%%%%%%%%
%
\begin{figure}
	\includegraphics[width=0.9\hsize]{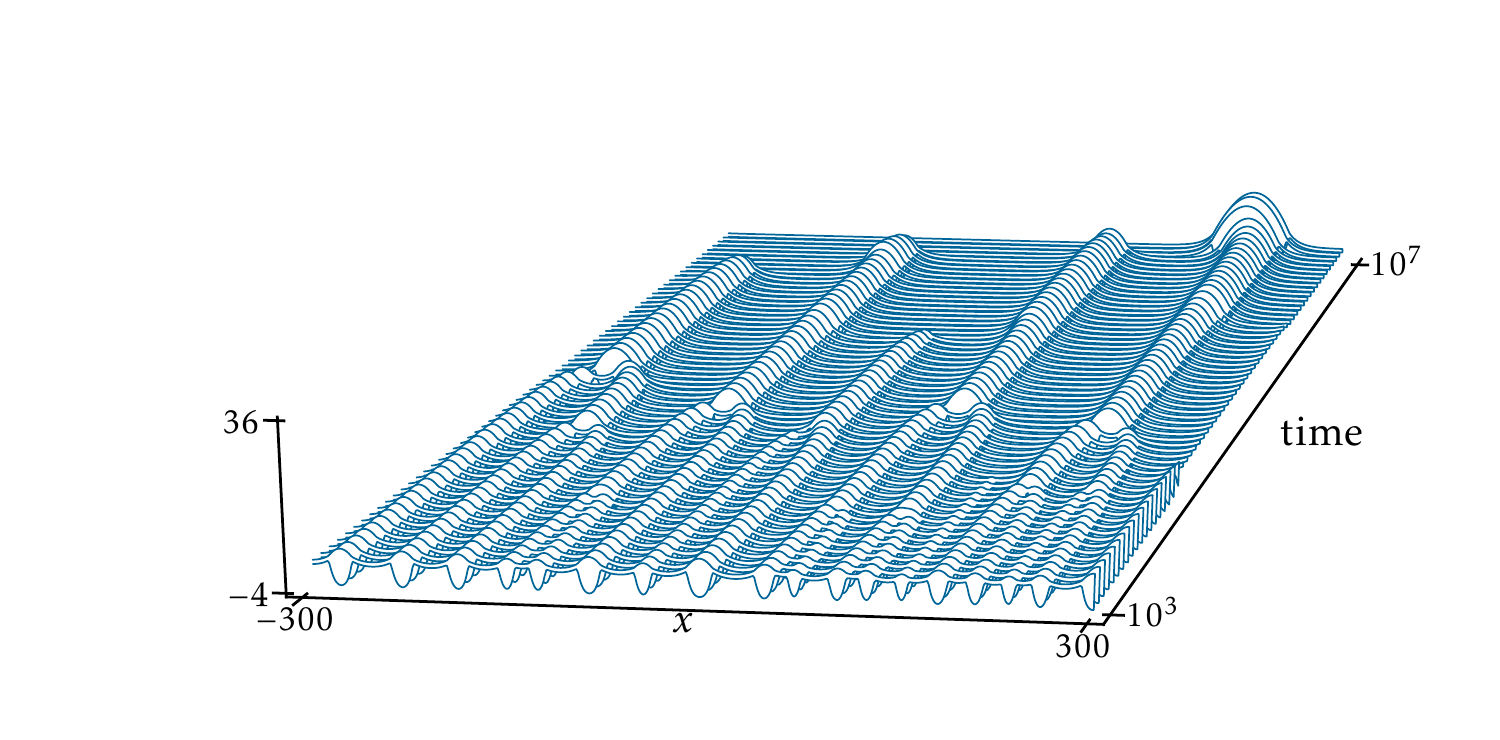}
	\caption{The space-time plot illustrates the coarsening of a droplet ensemble initially consisting of $N\approx200$ drops of partially wetting liquid on a substrate of intermediate softness $s=10^2$ [parameters as in Fig.~\ref{fig:twodrop-profiles}~(b)]. The domain size is $D=5000$, i.e., only a part is shown. The remaining parameters are as in Fig.~\ref{fig:spreading-single}. A corresponding time evolution can be seen in video~8 of the Supplementary Material.
	} 
	\label{fig:multidrop-profiles}
\end{figure}

\begin{figure}
	\includegraphics[width=0.8\hsize]{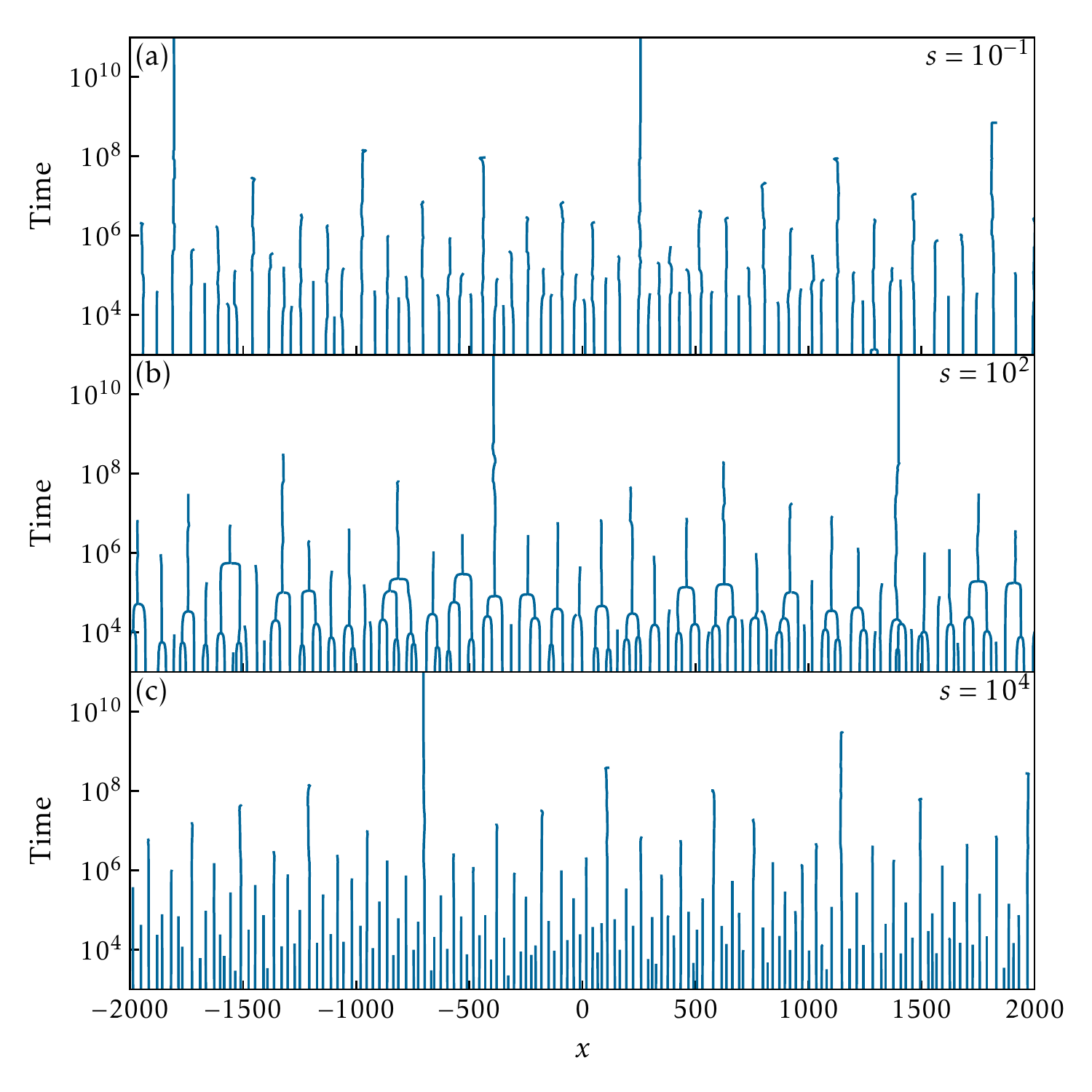}
	\caption{Space-time plots of the positions of all drop maxima on a logarithmic time scale for (a) the rigid limit at $s=10^{-1}$ (fewer initial drops and coarsening mainly via volume transfer, slight migrations are visible), (b) intermediate softness at $s=10^2$ (more initial drops, translation mode first dominates before the volume transfer mode prevails at larger drop size and distance), and (c) the soft limit at $s=10^4$ (further increased initial drop number, dominating volume mode, no visible drop migration). The domain size is $D=5000$, i.e., only part is shown. The remaining parameters are as in Fig.~\ref{fig:spreading-single}.}
	\label{fig:maxlines_multi}
\end{figure}

\begin{figure}
	\includegraphics[width=0.8\hsize]{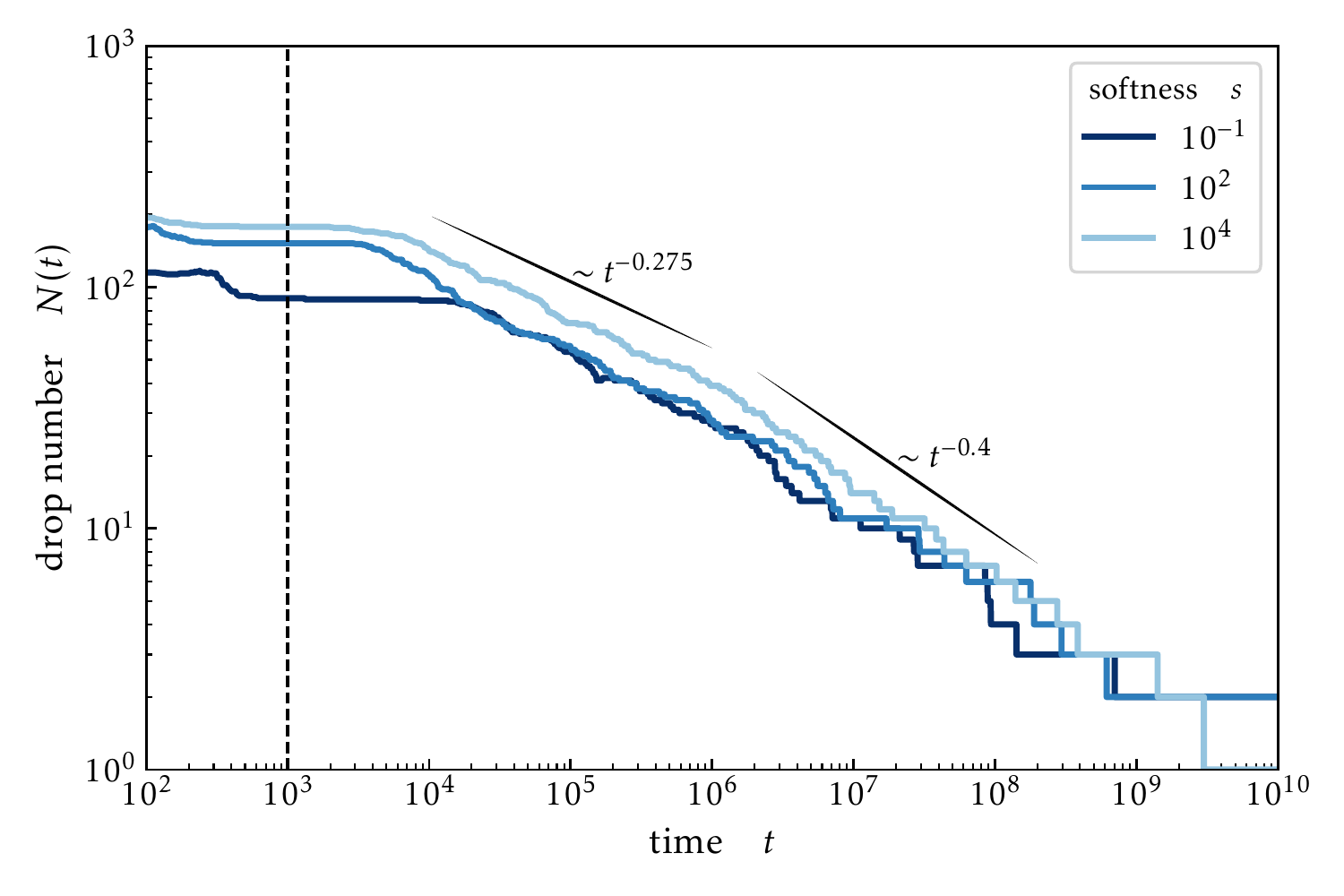}
	\caption{Dependence of drop number on time on a doubly logarithmic scale for softness $s$ as given in the legend. The initial stage of drop formation (dewetting) is finished at about $t=10^3$ (indicated by vertical line). Coarsening sets in shortly after when the respective plateaus end.}
	\label{fig:multidrop-number}
\end{figure} 

\begin{figure}
	\includegraphics[width=0.8\hsize]{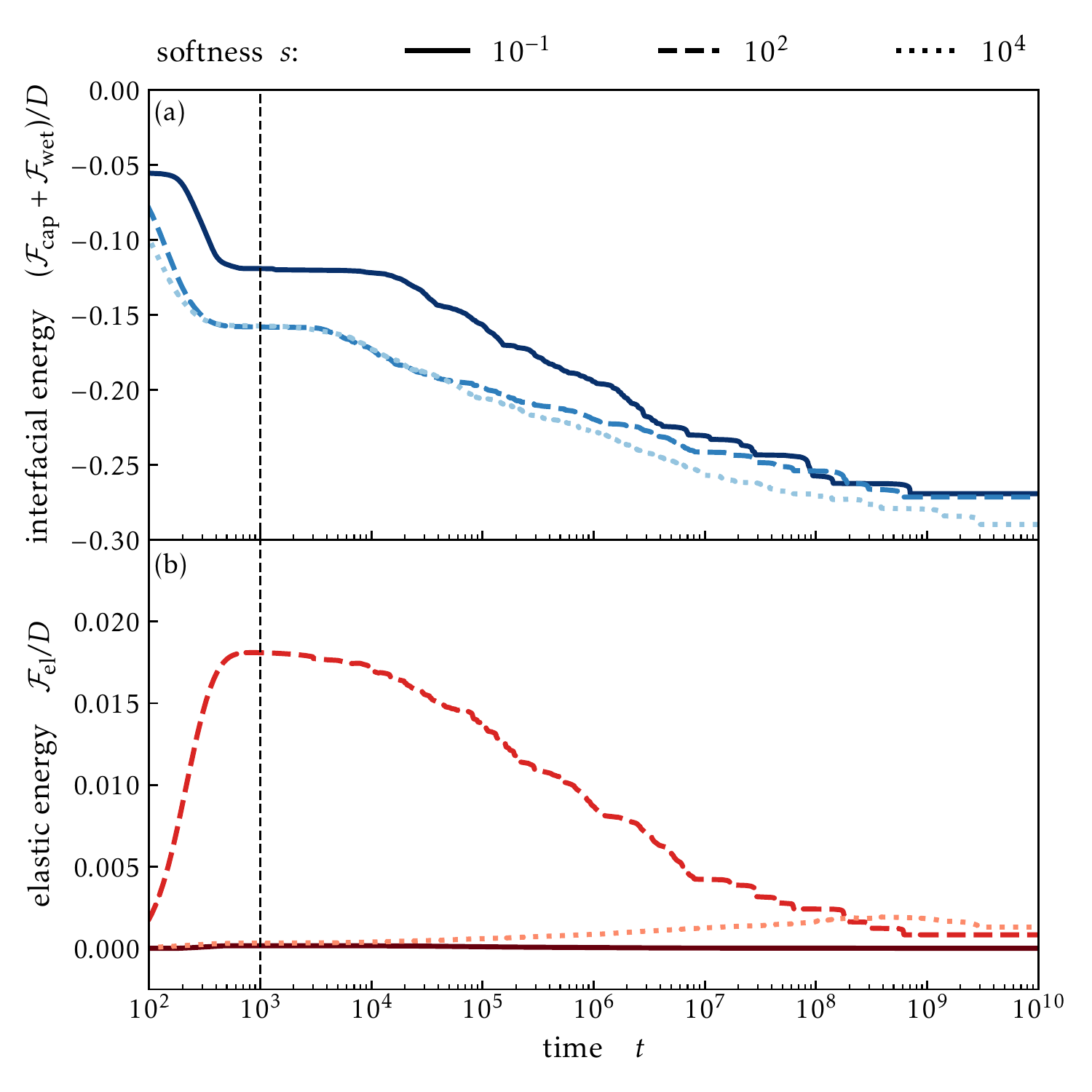}
	\caption{Dependence of (a) interface and (b) elastic mean energy density on time for softnesses $s$ as given in the legend. The strong decrease in interface energy for $10^2\leq t\leq10^3$ corresponds to the initial spinodal dewetting. After the subsequent plateau marked by the vertical dashed line coarsening sets in. Note that the contribution of elastic energy is much smaller than the one of the interface energy one, and almost negligible outside the range of intermediate elasticity, i.e.\ here $s=10^2$.}
	\label{fig:multidrop-energies}
\end{figure} 

Finally, we show that the developed model is also capable of describing the coarsening dynamics of large drop ensembles. Starting with a flat film of height $h_0=3$ on a rather large domain ($D=5000$), a small added noise initiates the early spinodal dewetting process \cite{ThVN2001prl,Wite2020am}. It produces a large ensemble of about 100-200 droplets which then undergo coarsening. Fig.~\ref{fig:multidrop-profiles} gives an exemplary space-time plot showing part of the domain and process for an ensemble of initially $\approx200$ drops on a substrate of intermediate softness ($s=10^2$). The subsequent Fig.~\ref{fig:maxlines_multi} shows space-time plots of the positions of all drop maxima for the same case of intermediate softness and, additionally, for the rigid and the liquid limit at $s=10^{-1}$ and $s=10^4$, respectively. The corresponding dependencies of drop number on time is presented in Fig.~\ref{fig:multidrop-number}.

Inspecting Fig.~\ref{fig:maxlines_multi}~(b) (also cf.~the magnification in Fig.~\ref{fig:multidrop-profiles}) shows that at intermediate softness nearly all coarsening events at early times occur by drop translation, i.e., two lines meet and join into one. In contrast, at later times from $t\approx10^4$ more and more drops vanish via volume transfer, i.e., single lines just terminate. Beyond $t\approx10^6$ no drop coalescence by translation is visible anymore. This agrees with the transition from dominant translation to volume transfer mode with increasing drop distance discussed in section~\ref{sec:coarse2} and described for rigid substrates in Ref.~\cite{GlWi2005pd}.

The mean drop volume also increases in time what should favor the translation mode. However, the effect of the accompanying increase in mean drop distance is stronger. Focusing at the rigid and liquid limit in Figs.~\ref{fig:maxlines_multi}~(a) and~(c), respectively, one notes that there coarsening by volume transfer dominates
in perfect agreement with the findings of section~\ref{sec:coarse2}. A closer look at Figs.~\ref{fig:maxlines_multi}~(a) even reveals the above discussed joint migration of neighbouring drops.

Next we focus on the evolution of the drop number in Fig.~\ref{fig:multidrop-number}. The figure includes earlier times than Fig.~\ref{fig:maxlines_multi} to indicate a plateau that clearly marks a time span when the initial spinodal dewetting process has ended and coarsening of the drop array has not yet begun.\footnote{Note that during the dewetting phase (not shown in Fig.~\ref{fig:multidrop-number}) our counting routine gives the number of maxima in the film profile, i.e., a mean wave number.} For the present parameters the plateaus in Fig.~\ref{fig:multidrop-number} range from $t\approx10^2$ to $t\approx10^4$, and represent an approximate unstable equilibrium. It is unstable with respect to combinations of the two-drop coarsening modes (discussed in section~\ref{sec:coarse2}), and various multi-drop modes; the latter, are, however, much weaker than the two-drop modes. Once coarsening starts at the end of the plateau, the drop number continuously decreases. Due to the discrete measure of the number of drops, individual steps in the curves are often visible and become prominent at late times when few drops remain. Qualitatively, the substrate softness does not seem to have an effect on long-time coarsening as the three curves are very similar. Empirically, a power law fit $N(t)\sim t^{-\alpha}$ gives approximately identical scaling exponents.  Based on the present low statistics one might even be tempted to distinguish two power laws. One in the range from $t\approx10^4$ to $t\approx10^6$ with $\alpha\approx0.275$ and one in the range from $t\approx10^6$ to $t\approx10^8$ with $\alpha\approx0.4$.

Finally, we briefly discuss how the various contributions to the total energy depend on time. Fig.~\ref{fig:multidrop-energies}~(a) gives the mean interface energy density $(\mathcal{F}_\mathrm{cap}+\mathcal{F}_\mathrm{wet})/D$ encompassing wetting and capillary contributions [see Eqs.~(\ref{eq:energy-cap}) and (\ref{eq:energy-wet})]. It monotonically decreases with time and decreases with increasing substrate softness. Fig.~\ref{fig:multidrop-energies}~(b) gives the mean elastic energy density $\mathcal{F}_\mathrm{el}/D$ [Eq.~\ref{eq:energy-elast}]. It is negligible in the rigid limit, small and nonmonotonic in the liquid limit, and monotonically decreases during the coarsening process (i.e., after the plateau) at intermediate softness. Even at  intermediate softness, where elasticity is most prominent, the elastic energy is much smaller than the interface energy. This indicates that coarsening is dominated by interface effects; this might offer an explanation why the coarsening exponent is not sensitive to the substrate softness.

%%%%%%%%%%%%%%%%%%%%%%%%%%%%%%%%%%%%%%%%%%%%%%%%%%%%%%%%%%%%%%%%%%%%%%%%%%%%%%%
\section{Summary and outlook}
\label{sec:conc}
%%%%%%%%%%%%%%%%%%%%%%%%%%%%%%%%%%%%%%%%%%%%%%%%%%%%%%%%%%%%%%%%%%%%%%%%%%%%%%%

We have developed a qualitative long-wave model that captures main features of the statics and dynamics of a liquid drop on an elastic substrate. The transparency and simplicity of the model rests in its gradient dynamics structure and the underlying free energy that reflects capillarity, wettability and compressional elasticity. The model has first been employed to investigate the double transition that occurs in the equilibrium contact angles when increasing the softness from an ideally rigid substrate towards a very soft, i.e., liquid substrate ~\cite{LWBD2014jfm}. Both, the dependency on softness and the scaling of the two transitions with drop volume given by our model, closely resemble the behaviour observed in Ref.~\cite{LWBD2014jfm} with a full elasticity model for drops of contact angle $\pi/2$ on thick incompressible elastic layers. In particular, the scaling of the typical softnesses where the two transitions occur with drop volume and the scalings of the wetting ridge height with softness in the different regimes agree very well. This very promising qualitative agreement should in the future be quantified by directly employing the full model for drops of small contact angles to determine effective parameters for the present long-wave model.

Second, we have investigated the spreading of a single drop of partially wetting liquid on elastic substrates. Dependencies of the dynamic contact angle on contact line velocity have been studied for different softnesses. Depending on the time scale ratio between elastic and liquid dynamics different scales could be introduced that allow to collapse sub-sets of curves. These rescalings are directly rooted in the  considerations in Refs.~\cite{KDGP2015nc,AnSn2020arfm} and indicates, that the dissipation in the substrate is dominant and increases with softness, i.e. that viscoelastic braking is present \cite{CaGS1996n,CaSh2001l}. Beyond that, the simplicity of our model allowed us to also investigate the case of completely wetting liquids. We find that viscoelastic braking is present in the completely wetting case as well, in contradiction to \cite{ChKu2020sm}. However, this effect is rather small and such is the shift between the curves for dynamic contact angles (Fig.~\ref{fig:spreading-wetting}) due to the less distinctive wetting ridge. 

Third, we have considered the coarsening dynamics of a pair of drops and of large ensembles of drops. In the case of two drops we have found that the dominant coarsening mode nonmonotonically changes when increasing the substrate softness. In both, the rigid and the soft limits, the volume transfer mode of coarsening (mass transfer mode, collapse mode, diffusion-controlled ripening, or Ostwald ripening) dominates while at intermediate softness the translation mode of coarsening (collision, coalescence, migration mode) dominates for sufficiently large drops. As such, the model recovers the ``inverted Cheerios effect", describing interaction of liquid drops mediated by elastic deformation of the solid substrate \cite{KPLW2016pnasusa} (in contrast to the Cheerios effect describing interaction of solid particles on a liquid substrate).  In the rigid limit the volume transfer mode is accompanied by migration of both drops into the direction of the smaller one in accordance with literature \cite{PiPo2004pf,GlWi2005pd}. Although, the nonmonotonic change in mode dominance has mainly been investigated by time simulations, it has furthermore be confirmed through a linear stability analysis for a closely related unstable steady two-drop constellation.

Based on the results on two-drop coarsening, finally, we have considered large drop ensembles and found that the change in dominance of the coarsening mechanisms can be found there as well. However, the strength of the translation mode decreases faster with increasing separation of the drops than the strength of the volume mode does. Hence, the translation dominated regime ends at a critical value as the drop number decreases and the mean distance between the drops become to large. As the initial drop contribution emerges from a dewetting film the number of drops at the beginning of the coarsening process is not controlled. Indeed the initial drop number was found to increase with increasing softness, which might be reasoned by the reduced width of drops that (partially) sank into the substrate. It will be interesting to apply the model in the future to investigate the coarsening of large ensembles of drops on a two-dimensional substrate.

      Beside of direct extensions of the presently studied examples, the presented model can be directly applied to a number of different situations, e.g., to study the sliding of individual drops under lateral driving, e.g., on an incline. Further,  the model lends itself easily to the investigation of moving contact lines on elastic substrates, e.g., in a Landau-Levich geometry \cite{SADF2007jfm,GTLT2014prl}. This allows one to investigate the appearance of stick-slip cycles as experimentally observed in \cite{KDNR2013sm,KBRD2014sm,GASK2018prl}. 
In its present form the model can be employed as an element of more complex models to study the interaction of complex fluids with soft substrates. For instance, it may be combined with thin-film models of active media \cite{LoEL2020sm,TSJT2020pre} or biofilms \cite{TJLT2017prl} to investigate the motion of active drops on soft substrates \cite{LoZA2014sm} and the growth of biofilms and cell aggregates on such substrates \cite{DoDB2012sm,CRAP2020e}.

Finally, we point out that the presented qualitative model may be expanded in a number of ways. One can explicitly introduce the effect of finite thickness on sliding \cite{ZDNL2018pnasusa}, which also enables to study phenomena such as durotaxis \cite{SCPW2013potnaos}. 
Furthermore, it will be interesting to see how the approach can be expanded to incorporate a description of the Shuttleworth effect \cite{AnSn2016el}, to account for substrate stretching  \cite{XuSD2018sm,SXHB2021prl}, or to systematically account for other rheologies and incompressibility of the elastic layer \cite{AnSn2020arfm}.
The latter should then allow to capture drop-drop repulsion and attraction via the inverted Cheerios effect \cite{KPLW2016pnasusa} not only the attractive case as the present model.

\acknowledgments

UT and JS acknowledges support by the Deutsche Forschungsgemeinschaft (DFG) via Grants TH781/12 and SN145/1-1, respectively within SPP~2171. We are grateful to Stefan Karpitschka and Johannes Kemper for discussions and calculations, respectively,  regarding an early version of the model. Further discussions with Linda Cummings, Lou Kondic and Simon Hartmann are acknowledged.

\appendix
\section{Elastic energy per area in the framework of linear elasticity}
\label{sec:app-Gkappav}

In this Appendix we discuss how elastic energy functionals can in principle be derived from the theory of linear elasticity. The central objects in elasticity theory are the  displacement vectors $\mathbf u$. When varying the displacement at the free surface, one performs a mechanical work that is equal to the stored elastic energy. Hence, the energy can in principle be expressed as a functional of $\mathbf u$ at the free surface. In the case of linear elasticity, this energy must be of quadratic form, 

\begin{equation}
\mathcal F_{\rm el} =  \frac{1}{2}G \int d^2\mathbf x \int d^2\mathbf x' \, \mathbf u(\mathbf x) \cdot \mathbf K(\mathbf x - \mathbf x') \cdot
 \mathbf u(\mathbf x').
\end{equation}
In this expression $G$ is the elastic (shear) modulus while $\mathbf K$ is a Green's function for an elastic layer that we assume to be homogeneous. 

We will now restrict ourselves to the case where the interface is shear free, and use that for small displacements  $u_z(\mathbf x) \simeq h(\mathbf x)$. Then, the elastic energy can be inferred from the interface shape 

\begin{equation}
\mathcal F_{\rm el} =  \frac{1}{2}G \int d^2\mathbf x \int d^2\mathbf x'  K(\mathbf x - \mathbf x') 
 h(\mathbf x)  h(\mathbf x'),
\end{equation}
where now $K$ is a scalar Green's function, relating normal displacement to normal stress. Explicit calculation of the Green's function requires solving the bulk-elastic problem inside the layer. In case the problem is invariant along $y$, the Green's function is known analytically in the form of a Fourier transform \cite{HANN1951tqjomaam,EPKV2021jofm}:

\begin{equation}
\widetilde K(q)^{-1}  = \frac{(1-\nu)}{q}\left[   \frac{(3-4\nu)\sinh(2qd) - 2qd}{(3-4\nu)\cosh(2qd) + 2(qd)^2 + 5 -12\nu +8\nu^2 }    \right]
\end{equation}
Here it is given as an inverse of the Green's function, which maps the traction to the displacement (while $\widetilde{K}(q)$ maps the displacement to the traction). It is of interest to consider some limits. In the short-wave limit ($qd \gg 1$), we find 

\begin{equation}
\widetilde K(q)^{-1} = \frac{1-\nu}{|q|}.
\end{equation}
In real-space this inverse Green's function is a logarithm (i.e. the surface displacement $h(x)$ due to a point force is logarithmic on a half-space). In the long-wave limit ($qd \ll 1$), the limiting behavior depends on the Poisson ratio. For $\nu \neq 1/2$, we find

\begin{equation}
\widetilde K(q)^{-1} = \frac{d(1-2\nu)}{2(1-\nu)}, 
\end{equation}
which is independent of $q$. This implies that the real-space Green's function is a $\delta$ function, i.e. 

\begin{equation}
K(x) =  \frac{2(1-\nu)}{d(1-2\nu)} \delta(x).
\end{equation}
We notice from this expression that there is a problem at $\nu=1/2$, so that an incompressible layer cannot deform under long-wave tractions. This property is a consequence of $\widetilde{K}(q=0)^{-1}=0$. 
Therefore, the long-wave expansion for incompressible media needs to go to the next order, to yield

\begin{equation}
\widetilde K(q)^{-1} =  \frac{d^3}{3} q^2, 
\end{equation}
which in real space implies

\begin{equation}\label{eq:incompressible}
h(x)= -\frac{d^3}{3G} \frac{\partial^2 p}{\partial x^2}. 
\end{equation}
Note the resemblance with the viscous thin-film response $h^3/(3\eta)\,\partial_{xx}p$: the Green's function in elasticity plays to role of the mobility in viscous fluids.  

We thus conclude that the assumption of a Green's function that is a $\delta$ function is rigorous in a specific limit: the long-wave limit $q d \ll 1$ for compressible layers ($\nu \neq 1/2$). In that case, the elastic energy reads

\begin{equation}
\mathcal F_{\rm el} = \frac{1}{2} \frac{2G(1-\nu)}{d(1-2\nu)} \int d^2 \mathbf x \, h(\mathbf x)^2,
\end{equation}
which is identical to Eq.~(\ref{eq:energy-elast}) in the main text. In this specific limit, we recover the connection

\begin{equation}
\kappa_v =  \frac{2G(1-\nu)}{d(1-2\nu)}.
\end{equation}
Notice once again that the formulation does not apply for incompressible layers ($\nu=1/2$). Another implicit assumption here is that we ignored any horizontal displacements that can be caused by shear stress induced by the fluid. In the considered long-wave limit, however, one can argue that shear stress is asymptotically small compared to the pressure.

%\bibliography{HeST2021}%

\begin{thebibliography}{100}

\bibitem{AlAu2021ijnme}
S.~Aland and P.~Auerbach.
\newblock A ternary phase-field model for wetting of soft elastic structures.
\newblock {\em Int. J. Numer. Methods Eng.}, 2021.
\newblock (online).
\newblock \href {http://dx.doi.org/10.1002/nme.6694}
  {\path{doi:10.1002/nme.6694}}.

\bibitem{AlMo2021ijnme}
S.~Aland and D.~Mokbel.
\newblock A unified numerical model for wetting of soft substrates.
\newblock {\em Int. J. Numer. Methods Eng.}, 122:903--918, 2021.
\newblock \href {http://dx.doi.org/10.1002/nme.6567}
  {\path{doi:10.1002/nme.6567}}.

\bibitem{AnSn2016el}
B.~Andreotti and J.~H. Snoeijer.
\newblock Soft wetting and the {S}huttleworth effect, at the crossroads between
  thermodynamics and mechanics.
\newblock {\em Europhys. Lett.}, 113:66001, 2016.
\newblock \href {http://dx.doi.org/10.1209/0295-5075/113/66001}
  {\path{doi:10.1209/0295-5075/113/66001}}.

\bibitem{AnSn2020arfm}
B.~Andreotti and J.~H. Snoeijer.
\newblock Statics and dynamics of soft wetting.
\newblock {\em Annu. Rev. Fluid Mech.}, 52:285--308, 2020.
\newblock \href {http://dx.doi.org/10.1146/annurev-fluid-010719-060147}
  {\path{doi:10.1146/annurev-fluid-010719-060147}}.

\bibitem{ABNP2002jfm}
C.~Andrieu, D.~A. Beysens, V.~S. Nikolayev, and Y.~Pomeau.
\newblock Coalescence of sessile drops.
\newblock {\em J. Fluid Mech.}, 453:427--438, 2002.
\newblock \href {http://dx.doi.org/10.1017/S0022112001007121}
  {\path{doi:10.1017/S0022112001007121}}.

\bibitem{BDHE2018prl}
D.~Baratian, R.~Dey, H.~Hoek, D.~van~den Ende, and F.~Mugele.
\newblock Breath figures under electrowetting: electrically controlled
  evolution of drop condensation patterns.
\newblock {\em Phys. Rev. Lett.}, 120:214502, 2018.
\newblock \href {http://dx.doi.org/10.1103/PhysRevLett.120.214502}
  {\path{doi:10.1103/PhysRevLett.120.214502}}.

\bibitem{BLHV2012prl}
J.~Blaschke, T.~Lapp, B.~Hof, and J.~Vollmer.
\newblock Breath figures: nucleation, growth, coalescence, and the size
  distribution of droplets.
\newblock {\em Phys. Rev. Lett.}, 109:068701, 2012.
\newblock \href {http://dx.doi.org/10.1103/PhysRevLett.109.068701}
  {\path{doi:10.1103/PhysRevLett.109.068701}}.

\bibitem{BCJP2013epje}
S.~Bommer, F.~Cartellier, S.~Jachalski, D.~Peschka, R.~Seemann, and B.~Wagner.
\newblock Droplets on liquids and their journey into equilibrium.
\newblock {\em Eur. Phys. J. E}, 36:87, 2013.
\newblock \href {http://dx.doi.org/10.1140/epje/i2013-13087-x}
  {\path{doi:10.1140/epje/i2013-13087-x}}.

\bibitem{BEIM2009rmp}
D.~Bonn, J.~Eggers, J.~Indekeu, J.~Meunier, and E.~Rolley.
\newblock Wetting and spreading.
\newblock {\em Rev. Mod. Phys.}, 81:739--805, 2009.
\newblock \href {http://dx.doi.org/10.1103/RevModPhys.81.739}
  {\path{doi:10.1103/RevModPhys.81.739}}.

\bibitem{Bormashenko2017}
E.Y. Bormashenko.
\newblock {\em Physics of Wetting: Phenomena and Applications of Fluids on
  Surfaces}.
\newblock De Gruyter, Berlin/Boston, 2017.

\bibitem{BoSD2014sm}
J.~B. Bostwick, M.~Shearer, and K.~E. Daniels.
\newblock Elastocapillary deformations on partially-wetting substrates: rival
  contact-line models.
\newblock {\em Soft Matter}, 10:7361--7369, 2014.
\newblock \href {http://dx.doi.org/10.1039/c4sm00891j}
  {\path{doi:10.1039/c4sm00891j}}.

\bibitem{CaGS1996n}
A.~Carr{\'e}, J.~C. Gastel, and M.~E.~R. Shanahan.
\newblock Viscoelastic effects in the spreading of liquids.
\newblock {\em Nature}, 379:432--434, 1996.
\newblock \href {http://dx.doi.org/10.1038/379432a0}
  {\path{doi:10.1038/379432a0}}.

\bibitem{CaSh2001l}
A.~Carr{\'e} and M.~E.~R. Shanahan.
\newblock Viscoelastic braking of a running drop.
\newblock {\em Langmuir}, 17:2982--2985, 2001.
\newblock \href {http://dx.doi.org/10.1021/la001600e}
  {\path{doi:10.1021/la001600e}}.

\bibitem{ChKu2020sm}
V.~Charitatos and S.~Kumar.
\newblock A thin-film model for droplet spreading on soft solid substrates.
\newblock {\em Soft Matter}, 16:8284--8298, 2020.
\newblock \href {http://dx.doi.org/10.1039/D0SM00643B}
  {\path{doi:10.1039/D0SM00643B}}.

\bibitem{CRAP2020e}
A.~Cont, T.~Rossy, Z.~Al-Mayyah, and A.~Persat.
\newblock Biofilms deform soft surfaces and disrupt epithelia.
\newblock {\em {eLife}}, 9:e56533, 2020.
\newblock \href {http://dx.doi.org/10.7554/elife.56533}
  {\path{doi:10.7554/elife.56533}}.

\bibitem{CrMa2009rmp}
R.~V. Craster and O.~K. Matar.
\newblock Dynamics and stability of thin liquid films.
\newblock {\em Rev. Mod. Phys.}, 81:1131--1198, 2009.
\newblock \href {http://dx.doi.org/10.1103/RevModPhys.81.1131}
  {\path{doi:10.1103/RevModPhys.81.1131}}.

\bibitem{Dai2010n}
S.~B. Dai.
\newblock On a mean field model for 1d thin film droplet coarsening.
\newblock {\em Nonlinearity}, 23:325--340, 2010.
\newblock \href {http://dx.doi.org/10.1088/0951-7715/23/2/006}
  {\path{doi:10.1088/0951-7715/23/2/006}}.

\bibitem{Genn1985rmp}
P.-G. de~Gennes.
\newblock Wetting: {S}tatics and dynamics.
\newblock {\em Rev. Mod. Phys.}, 57:827--863, 1985.
\newblock \href {http://dx.doi.org/10.1103/RevModPhys.57.827}
  {\path{doi:10.1103/RevModPhys.57.827}}.

\bibitem{GennesBrochard-WyartQuere2004}
P.-G. de~Gennes, F.~Brochard-Wyart, and D.~Qu{\'e}r{\'e}.
\newblock {\em Capillarity and wetting phenomena: Drops, bubbles, pearls,
  waves}.
\newblock Springer, New York, 2004.

\bibitem{DWCD2014ccp}
H.~A. Dijkstra, F.~W. Wubs, A.~K. Cliffe, E.~Doedel, I.~F. Dragomirescu,
  B.~Eckhardt, A.~Y. Gelfgat, A.~Hazel, V.~Lucarini, A.~G. Salinger, E.~T.
  Phipps, J.~Sanchez-Umbria, H.~Schuttelaars, L.~S. Tuckerman, and U.~Thiele.
\newblock Numerical bifurcation methods and their application to fluid
  dynamics: {A}nalysis beyond simulation.
\newblock {\em Commun. Comput. Phys.}, 15:1--45, 2014.
\newblock \href {http://dx.doi.org/10.4208/cicp.240912.180613a}
  {\path{doi:10.4208/cicp.240912.180613a}}.

\bibitem{DoKK1991ijbcb}
E.~Doedel, H.~B. Keller, and J.~P. Kernevez.
\newblock Numerical analysis and control of bifurcation problems {(II)
  B}ifurcation in infinite dimensions.
\newblock {\em Int. J. Bifurcation Chaos}, 1:745--72, 1991.
\newblock \href {http://dx.doi.org/10.1142/S0218127491000555}
  {\path{doi:10.1142/S0218127491000555}}.

\bibitem{DoDB2012sm}
S.~Douezan, J.~Dumond, and F.~Brochard-Wyart.
\newblock Wetting transitions of cellular aggregates induced by substrate
  rigidity.
\newblock {\em Soft Matter}, 8:4578--4583, 2012.
\newblock \href {http://dx.doi.org/10.1039/c2sm07418d}
  {\path{doi:10.1039/c2sm07418d}}.

\bibitem{Duss1979arfm}
E.~B. Dussan.
\newblock On the spreading of liquids on solid surfaces: {S}tatic and dynamic
  contact lines.
\newblock {\em Ann. Rev. Fluid Mech.}, 11:371--400, 1979.
\newblock \href {http://dx.doi.org/10.1146/annurev.fl.11.010179.002103}
  {\path{doi:10.1146/annurev.fl.11.010179.002103}}.

\bibitem{EGUW2019springer}
S.~Engelnkemper, S.V. Gurevich, H.~Uecker, D.~Wetzel, and U.~Thiele.
\newblock Continuation for thin film hydrodynamics and related scalar problems.
\newblock In A.~Gelfgat, editor, {\em Computational Modeling of Bifurcations
  and Instabilities in Fluid Mechanics}, Computational Methods in Applied
  Sciences, vol 50, pages 459--501. Springer, 2019.
\newblock \href {http://dx.doi.org/10.1007/978-3-319-91494-7_13}
  {\path{doi:10.1007/978-3-319-91494-7_13}}.

\bibitem{EnTh2019el}
S.~Engelnkemper and U.~Thiele.
\newblock The collective behaviour of ensembles of condensing liquid drops on
  heterogeneous inclined substrates.
\newblock {\em Europhys. Lett.}, 127:54002, 2019.
\newblock \href {http://dx.doi.org/10.1209/0295-5075/127/54002}
  {\path{doi:10.1209/0295-5075/127/54002}}.

\bibitem{EWGT2016prf}
S.~Engelnkemper, M.~Wilczek, S.~V. Gurevich, and U.~Thiele.
\newblock Morphological transitions of sliding drops - dynamics and
  bifurcations.
\newblock {\em Phys. Rev. Fluids}, 1:073901, 2016.
\newblock \href {http://dx.doi.org/10.1103/PhysRevFluids.1.073901}
  {\path{doi:10.1103/PhysRevFluids.1.073901}}.

\bibitem{EPKV2021jofm}
M.~H. Essink, A.~Pandey, S.~Karpitschka, C.~H. Venner, and J.~H. Snoeijer.
\newblock Regimes of soft lubrication.
\newblock {\em J. Fluid Mech.}, 915:A49, 2021.
\newblock (online).
\newblock \href {http://dx.doi.org/10.1017/jfm.2021.96}
  {\path{doi:10.1017/jfm.2021.96}}.

\bibitem{ExKu1996jcis}
C.~W. Extrand and Y.~Kumagai.
\newblock Contact angles and hysteresis on soft surfaces.
\newblock {\em J. Colloid Interface Sci.}, 184:191--200, 1996.
\newblock \href {http://dx.doi.org/10.1006/jcis.1996.0611}
  {\path{doi:10.1006/jcis.1996.0611}}.

\bibitem{FrKB1991pra}
D.~Fritter, C.~M. Knobler, and D.~A. Beysens.
\newblock Experiments and simulation of the growth of droplets on a surface
  (breath figures).
\newblock {\em Phys. Rev. A}, 43:2858--2869, 1991.
\newblock \href {http://dx.doi.org/10.1103/PhysRevA.43.2858}
  {\path{doi:10.1103/PhysRevA.43.2858}}.

\bibitem{GTLT2014prl}
M.~Galvagno, D.~Tseluiko, H.~Lopez, and U.~Thiele.
\newblock Continuous and discontinuous dynamic unbinding transitions in drawn
  film flow.
\newblock {\em Phys. Rev. Lett.}, 112:137803, 2014.
\newblock \href {http://dx.doi.org/10.1103/PhysRevLett.112.137803}
  {\path{doi:10.1103/PhysRevLett.112.137803}}.

\bibitem{GLBG2017csaea}
M.~Gielok, M.~Lopes, E.~Bonaccurso, and T.~Gambaryan-Roisman.
\newblock Droplet on an elastic substrate: Finite element method coupled with
  lubrication approximation.
\newblock {\em Colloid Surf. A-Physicochem. Eng. Asp.}, 521:13--21, 2017.
\newblock \href {http://dx.doi.org/10.1016/j.colsurfa.2016.08.001}
  {\path{doi:10.1016/j.colsurfa.2016.08.001}}.

\bibitem{GORS2009ejam}
K.~Glasner, F.~Otto, T.~Rump, and D.~Slepcev.
\newblock Ostwald ripening of droplets: the role of migration.
\newblock {\em Eur. J. Appl. Math.}, 20:1--67, 2009.
\newblock \href {http://dx.doi.org/10.1017/S0956792508007559}
  {\path{doi:10.1017/S0956792508007559}}.

\bibitem{Glas2008sjam}
K.~B. Glasner.
\newblock Ostwald ripening in thin film equations.
\newblock {\em SIAM J. Appl. Math.}, 69:473--493, 2008.
\newblock \href {http://dx.doi.org/10.1137/080713732}
  {\path{doi:10.1137/080713732}}.

\bibitem{GlWi2003pre}
K.~B. Glasner and T.~P. Witelski.
\newblock Coarsening dynamics of dewetting films.
\newblock {\em Phys. Rev. E}, 67:016302, 2003.
\newblock \href {http://dx.doi.org/10.1103/PhysRevE.67.016302}
  {\path{doi:10.1103/PhysRevE.67.016302}}.

\bibitem{GlWi2005pd}
K.~B. Glasner and T.~P. Witelski.
\newblock Collision versus collapse of droplets in coarsening of dewetting thin
  films.
\newblock {\em Physica D}, 209:80--104, 2005.

\bibitem{GrWi2009pd}
M.~B. Gratton and T.~P. Witelski.
\newblock Transient and self-similar dynamics in thin film coarsening.
\newblock {\em Physica D}, 238:2380--2394, 2009.
\newblock \href {http://dx.doi.org/10.1016/j.physd.2009.09.015}
  {\path{doi:10.1016/j.physd.2009.09.015}}.

\bibitem{HANN1951tqjomaam}
M.~Hannah.
\newblock Contact stress and deformation in a thin elastic layer.
\newblock {\em Q. J. Mech. Appl. Math.}, 4:94--105, 1951.
\newblock \href {http://dx.doi.org/10.1093/qjmam/4.1.94}
  {\path{doi:10.1093/qjmam/4.1.94}}.

\bibitem{HeHa2006}
M.~Heil and A.~L. Hazel.
\newblock Oomph-lib - an object-oriented multi-physics finite-element library.
\newblock In H.-J. Bungartz and M.~Sch{\"a}fer, editors, {\em Fluid-Structure
  Interaction: Modelling, Simulation, Optimisation}, pages 19--49. Springer,
  Berlin, Heidelberg, 2006.
\newblock \href {http://dx.doi.org/10.1007/3-540-34596-5_2}
  {\path{doi:10.1007/3-540-34596-5_2}}.

\bibitem{HACN2017sm}
A.~Hourlier-Fargette, A.~Antkowiak, A.~Chateauminois, and S.~Neukirch.
\newblock Role of uncrosslinked chains in droplets dynamics on silicone
  elastomers.
\newblock {\em Soft Matter}, 13:3484--3491, 2017.
\newblock \href {http://dx.doi.org/10.1039/c7sm00447h}
  {\path{doi:10.1039/c7sm00447h}}.

\bibitem{JXWD2011prl}
E.~R. Jerison, Y.~Xu, L.~A. Wilen, and E.~R. Dufresne.
\newblock Deformation of an elastic substrate by a three-phase contact line.
\newblock {\em Phys. Rev. Lett.}, 106:186103, 2011.
\newblock \href {http://dx.doi.org/10.1103/PhysRevLett.106.186103}
  {\path{doi:10.1103/PhysRevLett.106.186103}}.

\bibitem{Johnson1987}
K.~L. Johnson.
\newblock {\em Contact Mechanics}.
\newblock Cambridge University Press, Cambridge, 1987.

\bibitem{KBRD2014sm}
T.~Kajiya, P.~Brunet, L.~Royon, A.~Daerr, M.~Receveur, and L.~Limat.
\newblock A liquid contact line receding on a soft gel surface: dip-coating
  geometry investigation.
\newblock {\em Soft Matter}, 10:8888--8895, 2014.
\newblock \href {http://dx.doi.org/10.1039/c4sm01609b}
  {\path{doi:10.1039/c4sm01609b}}.

\bibitem{KDNR2013sm}
T.~Kajiya, A.~Daerr, T.~Narita, L.~Royon, F.~Lequeux, and L.~Limat.
\newblock Advancing liquid contact line on visco-elastic gel substrates:
  stick-slip vs. continuous motions.
\newblock {\em Soft Matter}, 9:454--461, 2013.
\newblock \href {http://dx.doi.org/10.1039/c2sm26714d}
  {\path{doi:10.1039/c2sm26714d}}.

\bibitem{KDGP2015nc}
S.~Karpitschka, S~Das, M.~van Gorcum, H.~Perrin, B.~Andreotti, and J.~H.
  Snoeijer.
\newblock Droplets move over viscoelastic substrates by surfing a ridge.
\newblock {\em Nat. Commun.}, 6:7891, 2015.
\newblock \href {http://dx.doi.org/10.1038/ncomms8891}
  {\path{doi:10.1038/ncomms8891}}.

\bibitem{KPLW2016pnasusa}
S.~Karpitschka, A.~Pandey, L.~A. Lubbers, J.~H. Weijs, L.~Botto, S.~Das,
  B.~Andreotti, and J.~H. Snoeijer.
\newblock Liquid drops attract or repel by the inverted {C}heerios effect.
\newblock {\em Proc. Natl. Acad. Sci. U. S. A.}, 113:7403--7407, 2016.
\newblock \href {http://dx.doi.org/10.1073/pnas.1601411113}
  {\path{doi:10.1073/pnas.1601411113}}.

\bibitem{KKCQ2017sm}
A.~Keiser, L.~Keiser, C.~Clanet, and D.~Quere.
\newblock Drop friction on liquid-infused materials.
\newblock {\em Soft Matter}, 13:6981--6987, 2017.
\newblock \href {http://dx.doi.org/10.1039/c7sm01226h}
  {\path{doi:10.1039/c7sm01226h}}.

\bibitem{Kita2014ejam}
G.~Kitavtsev.
\newblock Coarsening rates for the dynamics of slipping droplets.
\newblock {\em Eur. J. Appl. Math.}, 25:83--115, 2014.
\newblock \href {http://dx.doi.org/10.1017/S0956792513000314}
  {\path{doi:10.1017/S0956792513000314}}.

\bibitem{KrauskopfOsingaGalan-Vioque2007}
B.~Krauskopf, H.~M. Osinga, and J~Galan-Vioque, editors.
\newblock {\em Numerical Continuation Methods for Dynamical Systems}.
\newblock Springer, Dordrecht, 2007.
\newblock \href {http://dx.doi.org/10.1007/978-1-4020-6356-5}
  {\path{doi:10.1007/978-1-4020-6356-5}}.

\bibitem{LCWD2018l}
H.~Y. Liang, Z.~Cao, Z.~L. Wang, and A.~V. Dobrynin.
\newblock Surface stresses and a force balance at a contact line.
\newblock {\em Langmuir}, 34:7497--7502, 2018.
\newblock \href {http://dx.doi.org/10.1021/acs.langmuir.8b01680}
  {\path{doi:10.1021/acs.langmuir.8b01680}}.

\bibitem{LiGr2003l}
R.~Limary and P.~F. Green.
\newblock Dynamics of droplets on the surface of a structured fluid film:
  {L}ate-stage coarsening.
\newblock {\em Langmuir}, 19:2419--2424, 2003.

\bibitem{Lima2012epje}
L.~Limat.
\newblock Straight contact lines on a soft, incompressible solid.
\newblock {\em Eur. Phys. J. E}, 35:134, 2012.
\newblock \href {http://dx.doi.org/10.1140/epje/i2012-12134-6}
  {\path{doi:10.1140/epje/i2012-12134-6}}.

\bibitem{LoZA2014sm}
J.~Lober, F.~Ziebert, and I.~S. Aranson.
\newblock Modeling crawling cell movement on soft engineered substrates.
\newblock {\em Soft Matter}, 10:1365--1373, 2014.
\newblock \href {http://dx.doi.org/10.1039/c3sm51597d}
  {\path{doi:10.1039/c3sm51597d}}.

\bibitem{LoEL2020sm}
A.~Loisy, J.~Eggers, and T.~B. Liverpool.
\newblock How many ways a cell can move: the modes of self-propulsion of an
  active drop.
\newblock {\em Soft Matter}, 16:3106--3124, 2020.
\newblock \href {http://dx.doi.org/10.1039/d0sm00070a}
  {\path{doi:10.1039/d0sm00070a}}.

\bibitem{LoAL1996lb}
D.~Long, A.~Ajdari, and L.~Leibler.
\newblock Static and dynamic wetting properties of thin rubber films.
\newblock {\em Langmuir}, 12:5221--5230, 1996.
\newblock \href {http://dx.doi.org/10.1021/la9604700}
  {\path{doi:10.1021/la9604700}}.

\bibitem{LWBD2014jfm}
L.~A. Lubbers, J.~H. Weijs, L.~Botto, S.~Das, B.~Andreotti, and J.~H. Snoeijer.
\newblock Drops on soft solids: free energy and double transition of contact
  angles.
\newblock {\em J. Fluid Mech.}, 747:R1, 2014.
\newblock \href {http://dx.doi.org/10.1017/jfm.2014.152}
  {\path{doi:10.1017/jfm.2014.152}}.

\bibitem{MDSA2012prl}
A.~Marchand, S.~Das, J.~H. Snoeijer, and B.~Andreotti.
\newblock Capillary pressure and contact line force on a soft solid.
\newblock {\em Phys. Rev. Lett.}, 108:094301, 2012.
\newblock \href {http://dx.doi.org/10.1103/PhysRevLett.108.094301}
  {\path{doi:10.1103/PhysRevLett.108.094301}}.

\bibitem{MaGK2005jcis}
O.~K. Matar, V.~Gkanis, and S.~Kumar.
\newblock Nonlinear evolution of thin liquid films dewetting near soft
  elastomeric layers.
\newblock {\em J. Colloid Interface Sci.}, 286:319--332, 2005.
\newblock \href {http://dx.doi.org/10.1016/j.jcis.2004.12.034}
  {\path{doi:10.1016/j.jcis.2004.12.034}}.

\bibitem{MuWW2005jem}
A.~M{\"u}nch, B.~Wagner, and T.~P. Witelski.
\newblock Lubrication models with small to large slip lengths.
\newblock {\em J. Eng. Math.}, 53:359--383, 2005.
\newblock \href {http://dx.doi.org/10.1007/s10665-005-9020-3}
  {\path{doi:10.1007/s10665-005-9020-3}}.

\bibitem{OrDB1997rmp}
A.~Oron, S.~H. Davis, and S.~G. Bankoff.
\newblock Long-scale evolution of thin liquid films.
\newblock {\em Rev. Mod. Phys.}, 69:931--980, 1997.
\newblock \href {http://dx.doi.org/10.1103/RevModPhys.69.931}
  {\path{doi:10.1103/RevModPhys.69.931}}.

\bibitem{PAKZ2020prx}
A.~Pandey, B.~Andreotti, S.~Karpitschka, G.~J. van Zwieten, E.~H. van
  Brummelen, and J.~H. Snoeijer.
\newblock Singular nature of the elastocapillary ridge.
\newblock {\em Phys. Rev. X}, 10:031067, 2020.
\newblock \href {http://dx.doi.org/10.1103/PhysRevX.10.031067}
  {\path{doi:10.1103/PhysRevX.10.031067}}.

\bibitem{PWLL2014nc}
S.~J. Park, B.~M. Weon, J.~S. Lee, J.~Lee, J.~Kim, and J.~H. Je.
\newblock Visualization of asymmetric wetting ridges on soft solids with x-ray
  microscopy.
\newblock {\em Nat. Commun.}, 5:4369, 2014.
\newblock \href {http://dx.doi.org/10.1038/ncomms5369}
  {\path{doi:10.1038/ncomms5369}}.

\bibitem{PiPo2004pf}
L.~M. Pismen and Y.~Pomeau.
\newblock Mobility and interactions of weakly nonwetting droplets.
\newblock {\em Phys. Fluids}, 16:2604--2612, 2004.

\bibitem{PODC2012jpm}
M.~N. Popescu, G.~Oshanin, S.~Dietrich, and A.~M. Cazabat.
\newblock Precursor films in wetting phenomena.
\newblock {\em J. Phys.: Condens. Matter}, 24:243102, 2012.
\newblock \href {http://dx.doi.org/10.1088/0953-8984/24/24/243102}
  {\path{doi:10.1088/0953-8984/24/24/243102}}.

\bibitem{PBMT2005jcp}
A.~Pototsky, M.~Bestehorn, D.~Merkt, and U.~Thiele.
\newblock Morphology changes in the evolution of liquid two-layer films.
\newblock {\em J. Chem. Phys.}, 122:224711, 2005.
\newblock \href {http://dx.doi.org/10.1063/1.1927512}
  {\path{doi:10.1063/1.1927512}}.

\bibitem{RNRH2014n}
T.~Rinkel, J.~Nordmann, A.~N. Raj, and M.~Haase.
\newblock Ostwald-ripening and particle size focussing of {sub-10{nm} NaYF}$_4$
  upconversion nanocrystals.
\newblock {\em Nanoscale}, 6:14523--14530, 2014.
\newblock \href {http://dx.doi.org/10.1039/c4nr03833a}
  {\path{doi:10.1039/c4nr03833a}}.

\bibitem{RoBr2019mmmas}
M.~S. Roudbari and E.~H. van Brummelen.
\newblock Binary-fluid-solid interaction based on the
  {N}avier-{S}tokes-{K}orteweg equations.
\newblock {\em Math. Models Meth. Appl. Sci.}, 29:995--1036, 2019.
\newblock \href {http://dx.doi.org/10.1142/S0218202519410069}
  {\path{doi:10.1142/S0218202519410069}}.

\bibitem{SaMa2015jfm}
T.~Salez and L.~Mahadevan.
\newblock Elastohydrodynamics of a sliding, spinning and sedimenting cylinder
  near a soft wall.
\newblock {\em J. Fluid Mech.}, 779:181--196, 2015.
\newblock \href {http://dx.doi.org/10.1017/jfm.2015.425}
  {\path{doi:10.1017/jfm.2015.425}}.

\bibitem{SXEH2015sm}
F.~Schellenberger, J.~Xie, N.~Encinas, A.~Hardy, M.~Klapper, P.~Papadopoulos,
  H.~J. Butt, and D.~Vollmer.
\newblock Direct observation of drops on slippery lubricant-infused surfaces.
\newblock {\em Soft Matter}, 11:7617--7626, 2015.
\newblock \href {http://dx.doi.org/10.1039/c5sm01809a}
  {\path{doi:10.1039/c5sm01809a}}.

\bibitem{STSR2018nc}
R.~D. Schulman, M.~Trejo, T.~Salez, E.~Raphael, and K.~Dalnoki-Veress.
\newblock Surface energy of strained amorphous solids.
\newblock {\em Nat. Commun.}, 9:982, 2018.
\newblock \href {http://dx.doi.org/10.1038/s41467-018-03346-1}
  {\path{doi:10.1038/s41467-018-03346-1}}.

\bibitem{SkMa2004prl}
J.~M. Skotheim and L.~Mahadevan.
\newblock Soft lubrication.
\newblock {\em Phys. Rev. Lett.}, 92:245509, 2004.
\newblock \href {http://dx.doi.org/10.1103/PhysRevLett.92.245509}
  {\path{doi:10.1103/PhysRevLett.92.245509}}.

\bibitem{SXHB2021prl}
K.~Smith-Mannschott, Q.~Xu, S.~Heyden, N.~Bain, J.~H. Snoeijer, E.~R. Dufresne,
  and R.~W. Style.
\newblock Droplets sit and slide anisotropically on soft, stretched substrates.
\newblock {\em Phys. Rev. Lett.}, 126:158004, 2021.
\newblock \href {http://dx.doi.org/10.1103/physrevlett.126.158004}
  {\path{doi:10.1103/physrevlett.126.158004}}.

\bibitem{SADF2007jfm}
J.~H. Snoeijer, B.~Andreotti, G.~Delon, and M.~Fermigier.
\newblock Relaxation of a dewetting contact line. {P}art 1. {A} full-scale
  hydrodynamic calculation.
\newblock {\em J. Fluid Mech.}, 579:63--83, 2007.
\newblock \href {http://dx.doi.org/10.1017/S0022112007005216}
  {\path{doi:10.1017/S0022112007005216}}.

\bibitem{StVe2009jpm}
V.~M. Starov and M.~G. Velarde.
\newblock Surface forces and wetting phenomena.
\newblock {\em J. Phys.-Condens. Matter}, 21:464121, 2009.
\newblock \href {http://dx.doi.org/10.1088/0953-8984/21/46/464121}
  {\path{doi:10.1088/0953-8984/21/46/464121}}.

\bibitem{StarovVelardeRadke2007}
V.~M. Starov, M.~G. Velarde, and C.~J. Radke.
\newblock {\em Wetting and spreading dynamics}, volume 138.
\newblock Taylor and Francis, Boca Raton, 2007.
\newblock \href {http://dx.doi.org/10.1201/9781420016178}
  {\path{doi:10.1201/9781420016178}}.

\bibitem{SCPW2013potnaos}
R.~W. Style, Y.~Che, S.~J. Park, B.~M. Weon, J.~H. Je, C.~Hyland, G.~K. German,
  M.~P. Power, L.~A. Wilen, J.~S. Wettlaufer, and E.~R. Dufresne.
\newblock Patterning droplets with durotaxis.
\newblock {\em Proc. Natl. Acad. Sci. U. S. A.}, 110:12541--12544, 2013.
\newblock \href {http://dx.doi.org/10.1073/pnas.1307122110}
  {\path{doi:10.1073/pnas.1307122110}}.

\bibitem{StDu2012sm}
R.~W. Style and E.~R. Dufresne.
\newblock Static wetting on deformable substrates, from liquids to soft solids.
\newblock {\em Soft Matter}, 8:7177--7184, 2012.
\newblock \href {http://dx.doi.org/10.1039/c2sm25540e}
  {\path{doi:10.1039/c2sm25540e}}.

\bibitem{SJHD2017arcmp}
R.~W. Style, A.~Jagota, C.~Y. Hui, and E.~R. Dufresne.
\newblock Elastocapillarity: Surface tension and the mechanics of soft solids.
\newblock {\em Annu. Rev. Condens. Matter Phys.}, 8:99--118, 2017.
\newblock \href {http://dx.doi.org/10.1146/annurev-conmatphys-031016-025326}
  {\path{doi:10.1146/annurev-conmatphys-031016-025326}}.

\bibitem{Tann1979jpd}
L.~H. Tanner.
\newblock The spreading of silicone oil drops on horizontal surfaces.
\newblock {\em J. Phys. D}, 12:1473--1484, 1979.
\newblock \href {http://dx.doi.org/10.1088/0022-3727/12/9/009}
  {\path{doi:10.1088/0022-3727/12/9/009}}.

\bibitem{Tayl1998acis}
P.~Taylor.
\newblock Ostwald ripening in emulsions.
\newblock {\em Adv. Colloid Interface Sci.}, 75:107--163, 1998.
\newblock \href {http://dx.doi.org/10.1016/S0001-8686(98)00035-9}
  {\path{doi:10.1016/S0001-8686(98)00035-9}}.

\bibitem{Thie2007chapter}
U.~Thiele.
\newblock Structure formation in thin liquid films.
\newblock In S.~Kalliadasis and U.~Thiele, editors, {\em Thin Films of Soft
  Matter}, pages 25--93. Springer Vienna, Vienna, 2007.
\newblock \href {http://dx.doi.org/10.1007/978-3-211-69808-2\_2}
  {\path{doi:10.1007/978-3-211-69808-2\_2}}.

\bibitem{Thie2010jpcm}
U.~Thiele.
\newblock Thin film evolution equations from (evaporating) dewetting liquid
  layers to epitaxial growth.
\newblock {\em J. Phys.: Condens. Matter}, 22:084019, 2010.
\newblock \href {http://dx.doi.org/10.1088/0953-8984/22/8/084019}
  {\path{doi:10.1088/0953-8984/22/8/084019}}.

\bibitem{Thie2018csa}
U.~Thiele.
\newblock Recent advances in and future challenges for mesoscopic hydrodynamic
  modelling of complex wetting.
\newblock {\em Colloids Surf. A}, 553:487--495, 2018.
\newblock \href {http://dx.doi.org/10.1016/j.colsurfa.2018.05.049}
  {\path{doi:10.1016/j.colsurfa.2018.05.049}}.

\bibitem{Thiele2021lectureCont}
U.~Thiele.
\newblock Lecture ``{I}ntroduction to numerical continuation'', {WWU
  M}{\"u}nster, winter term 2020/21.
\newblock Video and Slides on zenodo, 2021.
\newblock \href {http://dx.doi.org/10.5281/zenodo.4544848}
  {\path{doi:10.5281/zenodo.4544848}}.

\bibitem{ThAP2016prf}
U.~Thiele, A.~J. Archer, and L.~M. Pismen.
\newblock Gradient dynamics models for liquid films with soluble surfactant.
\newblock {\em Phys. Rev. Fluids}, 1:083903, 2016.
\newblock \href {http://dx.doi.org/10.1103/PhysRevFluids.1.083903}
  {\path{doi:10.1103/PhysRevFluids.1.083903}}.

\bibitem{TBBB2003epje}
U.~Thiele, L.~Brusch, M.~Bestehorn, and M.~B{\"a}r.
\newblock Modelling thin-film dewetting on structured substrates and templates:
  {B}ifurcation analysis and numerical simulations.
\newblock {\em Eur. Phys. J. E}, 11:255--271, 2003.
\newblock \href {http://dx.doi.org/10.1140/epje/i2003-10019-5}
  {\path{doi:10.1140/epje/i2003-10019-5}}.

\bibitem{ThVN2001prl}
U.~Thiele, M.~G. Velarde, and K.~Neuffer.
\newblock Dewetting: {F}ilm rupture by nucleation in the spinodal regime.
\newblock {\em Phys. Rev. Lett.}, 87:016104, 2001.
\newblock \href {http://dx.doi.org/10.1103/PhysRevLett.87.016104}
  {\path{doi:10.1103/PhysRevLett.87.016104}}.

\bibitem{TVNB2001pre}
U.~Thiele, M.~G. Velarde, K.~Neuffer, M.~Bestehorn, and Y.~Pomeau.
\newblock Sliding drops in the diffuse interface model coupled to
  hydrodynamics.
\newblock {\em Phys. Rev. E}, 64:061601, 2001.
\newblock \href {http://dx.doi.org/10.1103/PhysRevE.64.061601}
  {\path{doi:10.1103/PhysRevE.64.061601}}.

\bibitem{TJLT2017prl}
S.~Trinschek, K.~John, S.~Lecuyer, and U.~Thiele.
\newblock Continuous vs. arrested spreading of biofilms at solid-gas interfaces
  - the role of surface forces.
\newblock {\em Phys. Rev. Lett.}, 119:078003, 2017.
\newblock \href {http://dx.doi.org/10.1103/PhysRevLett.119.078003}
  {\path{doi:10.1103/PhysRevLett.119.078003}}.

\bibitem{TSJT2020pre}
S.~Trinschek, F.~Stegemerten, K.~John, and U.~Thiele.
\newblock Thin-film modelling of resting and moving active droplets.
\newblock {\em Phys. Rev. E}, 101:062802, 2020.
\newblock \href {http://dx.doi.org/10.1103/PhysRevE.101.062802}
  {\path{doi:10.1103/PhysRevE.101.062802}}.

\bibitem{UeWR2014nmma}
H.~Uecker, D.~Wetzel, and J.~D.~M. Rademacher.
\newblock {pde2path} - a {Matlab} package for continuation and bifurcation in
  {2D} elliptic systems.
\newblock {\em Numer. Math.-Theory Methods Appl.}, 7:58--106, 2014.
\newblock \href {http://dx.doi.org/10.4208/nmtma.2014.1231nm}
  {\path{doi:10.4208/nmtma.2014.1231nm}}.

\bibitem{BRSZ2017}
E.~H. van Brummelen, M.~S. Roudbari, G.~Simsek, and K.~G. van~der Zee.
\newblock Binary-fluid--solid interaction based on the
  {N}avier-{S}tokes-{C}ahn-{H}illiard equations.
\newblock In S.~Frei, B.~Holm, T.~Richter, T.~Wick, and H.~Yang, editors, {\em
  Fluid-Structure Interaction}, pages 283--328. De Gruyter, Berlin, 2017.
\newblock \href {http://dx.doi.org/10.1515/9783110494259-008}
  {\path{doi:10.1515/9783110494259-008}}.

\bibitem{GASK2018prl}
M.~van Gorcum, B.~Andreotti, J.~H. Snoeijer, and S.~Karpitschka.
\newblock Dynamic solid surface tension causes droplet pinning and depinning.
\newblock {\em Phys. Rev. Lett.}, 121:208003, 2018.
\newblock \href {http://dx.doi.org/10.1103/PhysRevLett.121.208003}
  {\path{doi:10.1103/PhysRevLett.121.208003}}.

\bibitem{GKAS2020sm}
M.~van Gorcum, S.~Karpitschka, B.~Andreotti, and J.~H. Snoeijer.
\newblock Spreading on viscoelastic solids: are contact angles selected by
  neumann's law?
\newblock {\em Soft Matter}, 16:1306--1322, 2020.
\newblock \href {http://dx.doi.org/10.1039/c9sm01453e}
  {\path{doi:10.1039/c9sm01453e}}.

\bibitem{VeGY2001s}
R.~D. Vengrenovich, Y.~V. Gudyma, and S.~V. Yarema.
\newblock Ostwald ripening of quantum-dot nanostructures.
\newblock {\em Semiconductors}, 35:1378--1382, 2001.
\newblock \href {http://dx.doi.org/10.1134/1.1427975}
  {\path{doi:10.1134/1.1427975}}.

\bibitem{WeAS2013sm}
J.~H. Weijs, B.~Andreotti, and J.~H. Snoeijer.
\newblock Elasto-capillarity at the nanoscale: on the coupling between
  elasticity and surface energy in soft solids.
\newblock {\em Soft Matter}, 9:8494--8503, 2013.
\newblock \href {http://dx.doi.org/10.1039/c3sm50861g}
  {\path{doi:10.1039/c3sm50861g}}.

\bibitem{Whit2003jcis}
L.~R. White.
\newblock The contact angle on an elastic substrate. 1. the role of disjoining
  pressure in the surface mechanics.
\newblock {\em J. Colloid Interface Sci.}, 258:82--96, 2003.
\newblock \href {http://dx.doi.org/10.1016/S0021-9797(02)00090-5}
  {\path{doi:10.1016/S0021-9797(02)00090-5}}.

\bibitem{WTEG2017prl}
M.~Wilczek, W.~Tewes, S.~Engelnkemper, S.~V. Gurevich, and U.~Thiele.
\newblock Sliding drops - ensemble statistics from single drop bifurcations.
\newblock {\em Phys. Rev. Lett.}, 119:204501, 2017.
\newblock \href {http://dx.doi.org/10.1103/PhysRevLett.119.204501}
  {\path{doi:10.1103/PhysRevLett.119.204501}}.

\bibitem{Wite2020am}
T.~P. Witelski.
\newblock Nonlinear dynamics of dewetting thin films.
\newblock {\em AIMS Math.}, 5:4229--4259, 2020.
\newblock \href {http://dx.doi.org/10.3934/math.2020270}
  {\path{doi:10.3934/math.2020270}}.

\bibitem{WLJH2018sm}
H.~B. Wu, Z.~Z. Liu, A.~Jagota, and C.~Y. Hui.
\newblock Effect of large deformation and surface stiffening on the
  transmission of a line load on a neo-hookean half space.
\newblock {\em Soft Matter}, 14:1847--1855, 2018.
\newblock \href {http://dx.doi.org/10.1039/c7sm02394d}
  {\path{doi:10.1039/c7sm02394d}}.

\bibitem{XJBS2017nc}
Q.~Xu, K.~E. Jensen, R.~Boltyanskiy, R.~Sarfati, R.~W. Style, and E.~R.
  Dufresne.
\newblock Direct measurement of strain-dependent solid surface stress.
\newblock {\em Nat. Commun.}, 8:555, 2017.
\newblock \href {http://dx.doi.org/10.1038/s41467-017-00636-y}
  {\path{doi:10.1038/s41467-017-00636-y}}.

\bibitem{XuSD2018sm}
Q.~Xu, R.~W. Style, and E.~R. Dufresne.
\newblock Surface elastic constants of a soft solid.
\newblock {\em Soft Matter}, 14:916--920, 2018.
\newblock \href {http://dx.doi.org/10.1039/c7sm02431b}
  {\path{doi:10.1039/c7sm02431b}}.

\bibitem{ZDNL2018pnasusa}
M.~H. Zhao, J.~Dervaux, T.~Narita, F.~Lequeux, L.~Limat, and M.~Roche.
\newblock Geometrical control of dissipation during the spreading of liquids on
  soft solids.
\newblock {\em Proc. Natl. Acad. Sci. U. S. A.}, 115:1748--1753, 2018.
\newblock \href {http://dx.doi.org/10.1073/pnas.1712562115}
  {\path{doi:10.1073/pnas.1712562115}}.

\end{thebibliography}

\end{document}